\documentclass[sigconf]{acmart}

\AtBeginDocument{%
  }

\acmISBN{978-1-4503-XXXX-X/2018/06}

\usepackage{booktabs} % use booktabs instead of ugly regular tables
\usepackage{graphicx} % more figure options
\RequirePackage[l2tabu, orthodox]{nag} % checks for common LaTeX errors
\usepackage{microtype} % better typesetting
\usepackage{hyperref}
\usepackage[capitalize]{cleveref}
\usepackage{amsthm}
\usepackage{float}
\usepackage{enumitem}
\usepackage[x11names]{xcolor}% \usepackage{dblfloatfix}
\usepackage[utf8]{inputenc} % lenient to utf-8 characters like smart quotes

\definecolor{specaction}{HTML}{D55E00} % vermillion
\definecolor{specname}{HTML}{0072B2}   % blue
\definecolor{specop}{HTML}{008A67}     % green
\definecolor{specpunct}{HTML}{666666}

\newcommand{\heading}[1]{\vspace{4pt}\noindent\textbf{#1}.\enspace}

\newcommand{\mybullet}[1]
{\par\textbf{#1}:\enspace}

\newcommand{\sys}{\textsc{Cling}}
\newcommand{\manual}{\textsc{Manual}}
\newcommand{\ai}{\textsc{End-to-End Agent}}

\graphicspath{{assets/}} % put all your figures in this folder

\begin{document}

% \title{Cling: Supporting Cross-Page Coordination Interactively, Incrementally, and In-Situ}

% \title{Cling: Supporting Cross-Page Information Coordination Interactively, Incrementally, and In-Situ}

\title{Cling: Supporting Cross-Page Coordination with Human-Directed, In-Situ, and Small-Step Automation}

\author{Zhicheng Huang}
\affiliation{%
  \institution{University of California San Diego}
  \city{La Jolla, California}
  \country{United States}
}
\email{zhichenghuang@ucsd.edu}

\author{Yining Cao}
\affiliation{%
  \institution{University of California San Diego}
  \city{La Jolla, California}
  \country{United States}
}
\email{rimacyn@ucsd.edu}

\author{Haijun Xia}
\affiliation{%
  \institution{University of California San Diego}
  \city{La Jolla, California}
  \country{United States}
}
\email{haijunxia@ucsd.edu}

% \affiliation{%

%%
\renewcommand{\shortauthors}{Huang, et al.}
%%
% Abstract.

\begin{abstract}
% \todo{needs update}
% Many web-based information tasks require coordination across separate pages as information, functionality, and constraints are inherently distributed. Such coordination is often inefficient, requiring manually switch contexts and transfer information across pages. We conceptualize each web page as a structured representation of the underlying data with a specific schema, and cross-page coordination, thus, can be viewed as the transfer of information across schemas. Based on this insight, we introduce  a technical pipeline that constructs reusable schemas of individual pages and identifies correspondences between them. As the users interact with the web content, relevant correspondences are surfaced, allowing them to select the desired ones, which the system then executes automatically. We instantiate this approach in a prototype system, \sys{}. Our technical evaluation shows that \sys{} accurately identifies useful coordination opportunities across pages of diverse domains. A user study further demonstrates that users can readily understand and leverage these correspondences to complete tasks. Finally, we discuss the broad implications of our approach for enabling interoperability beyond the web.

Web-based information work spans independently developed pages, with goals and coordination needs often emerging through browsing.
Manual coordination requires repeated low-level actions, yet explicitly specifying operations for automation can interrupt engagement with information that shapes users' next steps.
Our approach constructs reusable, schema-based page representations and identifies coordination actions through correspondences across schemas.
We instantiate this approach in \sys{}, which surfaces coordination opportunities in situ for users to select, configure, and execute incrementally while directing the broader task.
A technical evaluation demonstrates accurate identification of useful opportunities across diverse domains.
A 12-participant study found that \sys{} reduced task durations by more than half compared with both manual interaction and an end-to-end AI agent, while requiring 61\% fewer clicks and 80\% fewer tab switches than manual interaction.
Its small-step automation supported control and verification as actions unfolded, while accommodating intent as it emerged and evolved through information encountered on the page.

\end{abstract}

% Many web-based information tasks require coordination across separate pages as information, functionality, and constraints are inherently distributed. Such coordination is often inefficient, requiring users to manually switch contexts and transfer information as needs emerge with the ongoing activity.

%    <concept>
%        <concept_id>10003120.10003121.10003129.10011756</concept_id>
%        <concept_desc>Human-centered computing~User interface programming</concept_desc>
%        <concept_significance>500</concept_significance>
%        </concept>
%    <concept>
%        <concept_id>10003120.10003121.10003126</concept_id>
%        <concept_desc>Human-centered computing~HCI theory, concepts and models</concept_desc>
%        <concept_significance>500</concept_significance>
%        </concept>
%  </ccs2012>

% \ccsdesc[500]{Human-centered computing~User interface programming}
% \ccsdesc[500]{Human-centered computing~HCI theory, concepts and models}

%%
% Keywords.
% \keywords{web sensemaking; web automation; cross-page coordination}

\begin{teaserfigure}
    \centering
    \includegraphics[width=0.94\textwidth]{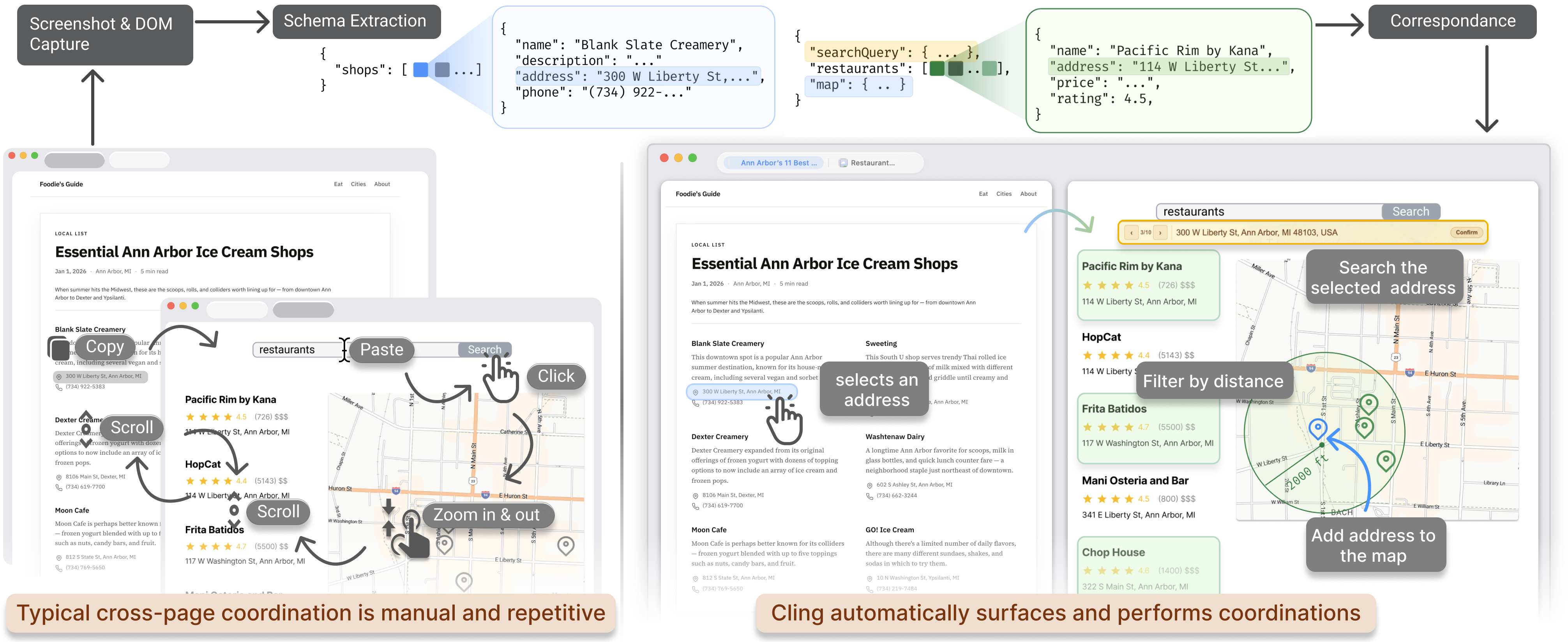}
    % \caption{Left: typical cross-page coordination is manual and repetitive. Right: \sys{} automatically surfaces coordination opportunities via schema correspondence and executes them in-situ.}
    \caption{\textbf{\sys{} Supports Cross-Page Coordination.} Left: typical cross-page coordination is manual and repetitive. Right: \sys{} automatically surfaces coordination opportunities via schema correspondence and executes them in-situ. Top: \sys{} extracts structured data models from pages and identifies cross-page correspondences to surface suggested coordination actions.}
    \label{fig:teaser}
    \Description[Overview of Cling]{Overview of Cling illustrated with an ice cream guide and a restaurant search page. On the left, arrows depict manual copying, pasting, clicking, scrolling, and zooming between separate browser windows. Across the top, a pipeline captures screenshots and DOMs, extracts structured schemas, and identifies correspondences among address fields, search inputs, and map elements. On the right, the pages appear side by side. An address selected in the guide is linked to a search field, highlighted nearby restaurants, and a map showing the selected location and a distance radius.}
\end{teaserfigure}

%% Teaser.
% \centering
%   \includegraphics[width=\textwidth]{figures/teaser.png}
%   \caption{Fairy Cursor Offers Active Assistance to keep user In-the-Flow}
%   \Description{A teaser figure divided into three rows. The first row illustrates a workflow using FairyCursor: the initial block shows a user typing a Slack query to locate the source of a screenshot, with FairyCursor providing real-time source suggestions; subsequent blocks depict how FairyCursor guides the user to the source website and scrolls to the correct position; and the final block demonstrates synchronized behavior—when the user hovers over a specific part of the website, the corresponding section of the video is highlighted. The second row presents the four types of in-flow assistance provided by FairyCursor, while the third row identifies the four common inefficiencies that disrupt the workflow.

%%
\maketitle

\section{Introduction}
\label{sec:introduction}

% many web tasks involve this pattern

% however, it's challenging to do these coordination patterns because XXX

% there are solutions: manual, and automation (alluding baselines). they each have their cost and benefits.

% we propose a solution (the automation paradigm) that xxx

% our approach, and conceptualization

% what did we do, motivating scenerios, what have we found

% structure of this paper, contributions

% Many web-based information tasks span multiple pages.
% People cross-reference, compare, search, or write across page boundaries.
% We refer to this class of interaction as \emph{cross-page coordination}, where one must coordinate multiple pages to accomplish a task. 

Web-based information work often spans many pages~\cite{hollan2000distributed}. When searching for an apartment before relocating, for example, a person may examine a rental listing, view its location on a map, browse its reviews on a community website, compare alternatives, and record selected details in a personal shortlist. In this process, the user needs to frequently transfer information across pages, cross-examine content on different pages, and integrate details from different pages. Such \emph{cross-page coordination} is a recurring pattern of everyday information work, yet it is often tedious and disruptive: users must mentally keep track of numerous disconnected pages while manually performing low-level navigation and information-transfer operations.

Supporting cross-page coordination, however, presents both structural and interaction challenges. Structurally, webpages are independently developed and generally lack incentives for interoperability. Shared data structures and protocols can support coordination, but typically only within the boundaries of integrated enterprise applications or ecosystems~\cite{googleworkspace, microsoft, notion}. Meanwhile, coordination often emerges opportunistically as work unfolds, and therefore, cannot be anticipated and supported with system features~\cite{contextual_facet}. A detail encountered on one page may prompt a new search, introduce a new comparison criterion, or shift the user's priorities. Cross-page coordination can therefore range from a one-off lookup to repeated information transfers, and from a step in a predefined workflow to part of an exploratory activity whose goals and constraints evolve throughout the process.

Client-side web automation tools~\cite{chasins2018rousillon, chen2023miwa, pu2023dilogics, pu2022semanticon, zhang2025multi} and recent computer-use agents~\cite{browseruse2025, operator2025, openclaw2025}
offer ways to execute operations across page boundaries. As these capabilities advance, an important interaction question is how to make desired automation available within ongoing, human-directed work. Specifying a workflow or delegating a task can be useful when the goals and constrains are already clear. In other situations, users discover what they want to do through inspecting information and making situated ~\cite{suchman1987plans}  and contextual judgments~\cite{marchionini2006exploratory,soufan2022searching,wildemuth2012assigning,white2008evaluating,kules2008creating}. They may need assistance with an individual lookup or compare corresponding fields while continuing to determine the next step themselves. Supporting this process requires making automation accessible at the moment a need becomes clear, while allowing users to continue directing the activity. Otherwise, users are left to either specify each operation for automation, interrupting their engagement with the participating pages and opportunities to discover their next steps, or to carry them out manually through repeated low-level actions.

We investigate how cross-page coordination can be supported by incrementally surfacing in-situ automation opportunities that users can selectively invoke as needs arise. Our approach uses information correspondences across pages as the basis for identifying potential coordination actions. For example, an address in a listing and a search field on a map page suggest a possible value transfer, while corresponding attributes across listings (e.g., square footage and price) suggest opportunities for cross-referencing or comparison.

Concretely, we model webpage content as a set of entities, each represented by entity schema, runtime values, and references to its original DOM elements. Under this representation, cross-page coordination can be expressed as operations for selecting, filtering, navigating, and transferring information among corresponding entities. Based on this formalization, we define five coordination patterns that capture distinct forms of cross-page correspondence. We use LLMs to construct the schematic representation of each webpage, while deliberately relying on rule-based correspondence detection to surface small-step automation opportunities in real time. As people interact with live webpages, the system dynamically surfaces coordination opportunities derived from cross-page entity correspondences.
Depending on their current goals, people can then selectively apply any automation opportunity in-situ to speed though tedious low-level manual operations while retaining control over the direction and progression of the broader task. We instantiate this approach in a protoype system, \sys{}.

We assess our approach through both technical and user evaluations. 
In the technical evaluation, 
we evaluated our technical pipeline~(\cref{sec:technical-pipeline}) across 48 webpages spanning 3 common web task domains. Two human coders annotated all 220 coordination opportunities surfaced by the pipeline, and rated them 92.5\% accurate with a mean usefulness of 4.48/5. On average, the page-model extraction took 5.82s across three independent runs.

We further conducted a two-part user study with 12 participants, examine \sys{} in both structured cross-page coordination and broader exploratory tasks.
In structured tasks, \sys{} combined the efficiency of end-to-end automation~(compared with \manual{}, it required 61.1\% fewer clicks, 99.6\% fewer copy/paste actions, and 80.2\% fewer tab switches, while completing tasks 2.07$\times$ as fast as \manual{} and 2.14$\times$ as \ai{}). 
At the same time, participants reported substantially greater control with \sys{} than with \ai{} (2.83 points higher on a 5-point scale, 95\% CI [2.33, 3.33], $p<0.001$), while reporting a level of control comparable to, and slightly higher than, \manual{}. Moreover, \sys{}'s in-situ and small-step automation allowed participants to inspect and correct mistakes as automation unfold, much as they did during manual work, while accommodating dynamic user intents that emerged and evolved over the course of the task.
% through information serendipitously encountered on the page.

% We conducted a two-part user study with 12 participants, examining \sys{} in structured cross-page coordination and broader exploratory tasks.
% \sys{} reduced manual effort and completion time while preserving participants' sense of control and supporting inspection as automation unfolded.
% In broader tasks, participants incorporated coordination into ongoing exploration, where in-situ information continually shaped and revised their criteria, preferences, and plans.

% We end by discussing the promises, limitations, and design implications of supporting cross-page coordination.
In summary, this paper makes the following contributions:
\begin{itemize}[leftmargin=*]
    \item The formalization of a class of interaction, which we call cross-page coordination as a set of 5 coordination patterns.
    \item A technical pipeline that constructs reusable entity-based page models and identifies coordination opportunities through page-model correspondence.
    \item User interfaces and interaction techniques for representing, surfacing, and invoking in-situ, small-step automation.
    \item \sys{}, a prototype system that instantiates these technical and interaction approaches.
    \item Technical and user evaluations that demonstrate the feasibility, effectiveness, and values of supporting cross-page coordination with human-directed, in-situ, and small-step automation
\end{itemize}

\section{Related Work}
\label{sec:related-work}
% \todo{last few bib}
% What the the several areas of research that relate to you work?
% Give a short description here of the related areas.
% Information tasks spanning multiple materials and systems often incur frequent context switching, information transfer, and coordination
% across interfaces. They have long been a central focus in HCI research. Prior work can be broadly categorized into four themes based on their approaches: integration, interoperability, automation, and correspondence-driven synchronization.
Information tasks spanning multiple materials and systems often
require context switching, information transfer, and coordination
across interfaces. Prior work has addressed these demands through
integrated environments, interoperability, task automation, and
correspondence-driven interaction. We review these approaches to situate \sys{}’s support for human-directed, in situ, small-step coordination across webpages.

\subsection{Integrated Workspaces for Distributed Information Tasks}
A common strategy for reducing coordination overhead is to build integrated environments for tasks that span multiple materials or processing steps, such as content creation \cite{adobecreativecloud, cao2025compositional, cao2023dataparticles}, data analysis \cite{microsoftfabric, jupyter, databricks, tableau}, programming \cite{mcilroy1978unix, visualstudiocode}, and task management \cite{reframingdesktop, googleworkspace, microsoft, visualstudiocode}. Although these systems promise an ``all-in-one'' experience, they typically demand substantial engineering effort and often depend on complex organizational infrastructure.

Integrated environments have also been used to support human--AI collaboration: Buschek et al. integrate diverse AI-powered tools into a unified writing environment \cite{CollageNewWriting}, and Cao et al. combine core video-editing structures---such as canvas, document, grid, and timeline---within a single environment to enable human--AI co-creation \cite{cao2025compositional}.
In such systems, coordination is often delegated to AI through predefined rules. However, building these environments has traditionally required developers to predefine information sources, structures, and tools.
Cao et al. recently reduced this burden with an approach that uses AI to dynamically generate environments in response to evolving user criteria~\cite{jelly}.

% Beyond intensive tasks that require dedicated environments, many everyday information activities unfold in situ, such as web browsing which this work aims to support. 

Browser-based workspaces address similar organizational needs across web materials.
Prior work has explored augmenting browsing environments with in-situ support. For example, systems have incorporated query suggestions~\cite{palani2021conotate} and note-taking~\cite{ward2021orgbox, crescenzi2019towards, liu2019unakite} into browsing workflows, transformed tabs into spatial or canvas-based workspaces to better scale browsing activity~\cite{jiang2025orca}, and externalized users' evolving interests through compositional filters and lenses~\cite{chang2019searchlens}. Other systems help users externalize mental models such as comparing across collected information~\cite{chang2020mesh, liu2021reuse, kuznetsov2022fuse, liu2022crystalline}, and triage during sensemaking~\cite{liu2022wigglite}. 

% Our work extends this direction by supporting coordination across heterogeneous web pages during browsing.

These approaches provide representations for organizing and reasoning across distributed materials. \sys{} complements this direction by supporting operations between the materials themselves: users can navigate corresponding entities, filter options, and transfer information across the live webpages involved in their work.

% \subsection{Interoperable Systems and Materials}
\subsection{Interoperability and End-User Composition}

Information materials and systems that support interoperability generally incur lower correspondence costs, as shared representations mitigate the need for manual transformation and coordination. 
Nelson’s concept of transclusion enables portions of one document to be embedded within another, facilitating the flexible composition of both information and functionality \cite{Transclusion}. This idea has been further extended in systems such as Webstrates \cite{klokmose2015webstrates} and subsequent work \cite{radle2017codestrates, klokmose2019videostrates, klokmose2024mywebstrates}, which support the sharing and synchronization of information and functionality across heterogeneous environments (e.g., web pages and devices). Beaudouin-Lafon et al. proposes a specification language to reason about structural analogies of interfaces~\cite{beaudouin2026belidor}. Interoperability has also been pursued through system-level mechanisms, including proprietary approaches such as Microsoft’s Object Linking and Embedding (OLE)~\cite{microsoftoleds} and Apple’s OpenDoc~\cite{Apple_OpenDoc, sleeter1996opendoc}.

End-user composition and customization allow people to adapt existing resources to their own needs. Hartmann et al.’s d.mix explores programming by sampling existing websites to create new web applications~\cite{hartmann2007programming}. Wildcard exposes application data through a spreadsheet interface, enabling users to customize website behavior~\cite{litt2020wildcard}. Together, these approaches show how existing data, functionality, and interface structures can become resources for composition. Min et al. explore malleable overview-detail interfaces that let users customize content, composition, and layout, including revealing attributes across views and using AI to transform or generate attributes~ \cite{min2025malleable}. These approaches make existing data, functionality, and interface structures available for user-directed adaptation and composition.

However, achieving interoperability typically requires the adoption of shared protocols, data schemas, and frameworks, which introduces substantial development overhead and limits practical adoption. In many cases, such efforts are further constrained within platform-specific ecosystems, resulting in ``walled gardens” that hinder integration across heterogeneous systems. More fundamentally, supporting interoperability remains intrinsically challenging, as developers cannot realistically anticipate the full diversity of user needs and usage contexts to define the scope of interoperability \cite{karger2007haystack}. Rather than relying on developers to conform to a single, globally agreed-upon standard, an assumption that is often untenable, we explore an alternative approach: leveraging generative AI to dynamically parse diverse web pages into shared, structured representations, thereby enabling on-demand interoperability across otherwise incompatible systems.

\subsection{Task Automation and In-Situ AI Assistance}
Traditional web automation systems, including web-scraping scripts~\cite{beautiful_soup, selenium_ide, cypress, puppeteer} and programming-by-demonstration approaches~\cite{barman2016ringer, chasins2018rousillon, chen2023miwa, dong2022webrobot, pu2023dilogics, pu2022semanticon}, offer reliable, performant ways to navigate websites and collect information at scale. Yet they generally assume well-defined environments and well-specified success criteria, assumptions that often fail in exploratory, open-ended, and uncertain tasks~\cite{white2009exploratory, white2008evaluating, wildemuth2012assigning, soufan2022searching}. Recent LLM-based agents~\cite{browseruse2025, operator2025, openclaw2025, liu2026reuseit} optimize for task completion by committing to browsing trajectories shaped by model priors and interface cues~\cite{gebreegziabher2026behavioral, sharma2024generative}. In contrast, many information tasks evolve incrementally, motivating in-situ assistance that users can invoke in small steps while deciding how to proceed.
% call for support that augments rather than replaces exploratory browsing.

Context-sensitive assistance has a long history in interface agents and mixed-initiative interaction, including anticipating browsing interests, inferring assistance needs, and deciding when to intervene~\cite{lieberman1995letizia, horvitz2013lumiere, horvitz1999principles}. Recent work revisits these questions: Fairy Cursor explores a cursor-based agent for in-the-flow assistance~\cite{cao2026exploring}, while Just-In-Time Objectives infers immediate goals from observed activity to specialize AI-generated tools and responses~\cite{lam2026just}. \sys{} builds on this direction by surfacing individual coordination actions as small-step automation that users can selectively execute within ongoing browsing.

% A related line of research is action generalization, which enables actions such as selection, text entry, and clicking to be applied across multiple structurally similar items
Action generalization provides a related precedent for this granularity of support. Prior techniques allow selection, text entry, and clicking to be applied across multiple structurally similar items
~\cite{landauer1995visual, miller2001interactive, miller2002multiple, miller2004cluster}. Zhang et al. recently brought this idea to the web, allowing users to perform the same action simultaneously on multiple selected targets~\cite{zhang2025multi}. However, such techniques rely on the involved pages having similar structures, such as in their CSS selectors or DOM placements, an assumption that breaks down for heterogeneous pages. Our approach addresses this limitation by lifting heterogeneous pages to their schemas, where cross-page coordination can be generalized despite structural differences.

% \subsection{Data-Driven Correspondence}
\subsection{Correspondence-Driven Interaction}
A substantial body of research in HCI and information visualization has investigated how semantic correspondences within content can be leveraged to enable coordinated and automated interaction. For example, systems built on time-aligned transcripts---such as QuickCut \cite{truong2016quickcut} and CrossCast \cite{crosscast}—exploit temporal alignment between text and video to support script-based editing workflows. Similarly, Data Particles \cite{cao2023dataparticles} and CrossData \cite{crossData} utilize the semantic correspondence between natural language, data operations, and visual encodings to facilitate the authoring of data-driven interactive articles with rich visualizations. In visualization, prior work has also explored leveraging shared data dimensions to support coordinated view management across multiple plots \cite{roberts2007state, boukhelifa2003coordination}. For web browsing specifically, Contextual Facets transforms webpage elements into controls for filtering and retrieving related pages~\cite{contextual_facet}. Dontcheva et al. enable users to establish relationships between websites, collect related content into personalized cards, and reuse these relationships through search templates~\cite{dontcheva2007relations}. 

By operationalizing these correspondences, such systems can automate a range of otherwise tedious low-level operations, including navigation, view coordination, content transfer, and transformation.
Our work aligns with this line of research in leveraging inherent correspondences within information artifacts. However, rather than relying on developer-specified mappings, we target open-ended web environments where such correspondences are not predefined. We introduce an approach that leverages generative AI to parse heterogeneous web pages into structured schemas and employs heuristic strategies to infer correspondences at interaction time, enabling real-time support for navigation and information transformation across heterogeneous pages.

\section{Example Use Case}
\label{sec:example-usage}

Polly, staff at a university, is co-organizing a panel on AI's impact on education.
She is responsible for identifying potential participants. The organizing team maintains a shared spreadsheet listing candidate invitees, including their name, affiliation, contact information, and a brief justification for inclusion.

To populate this spreadsheet, Polly needs to visit faculty directories from different departments, looks up their profiles on Google Scholar, reviews their publications, and decides whether they are a good fit. In a typical workflow, for each faculty member, she first checks whether they are already in the spreadsheet by copying their name and searching for it (e.g., using \texttt{cmd+F}) in another tab. If not found, she needs to then switch to Google Scholar, copies the name into the search box, submits the query, and reviews the profile. When appropriate, Polly needs to manually copy the name, affiliation, and email back into the spreadsheet, carefully aligning each field with the correct column in the spreadsheet. If the faculty member is already listed, Polly still needs to verify whether the contact information is up to date, requiring her to locate the corresponding row and cross-reference multiple sources. She repeats this process for each faculty member.

With \sys{}, Polly first places the faculty directory, Google Scholar, and the spreadsheet side by side. When Polly selects a faculty name on the directory page, \sys{} indicates whether the person is already in the spreadsheet. If not, \sys{} suggests transferring the selected name into the Google Scholar search field and submitting the query. After reviewing the publications, Polly decides to include the faculty member and selects ``Add to spreadsheet.'' \sys{} then aggregates and reformats the relevant information into a new row matching the schema. After previewing the entry, Polly confirms the insertion. If the faculty member is already in the spreadsheet, \sys{} surfaces a ``Go to'' action that navigates directly to the corresponding row. Polly can then quickly verify and update contact information by cross-referencing the directory and Google Scholar pages, without manually searching or aligning fields. Using \sys{}'s navigation widget, Polly can step to the next researcher while preserving the coordination context across the pages. At each step, \sys{} updates the relevant actions---for example, whether the researcher is already in the spreadsheet, where their row is, and by issuing the next Google Scholar search automatically. Polly can thus move continuously through researchers without repeating the same search, tab-switching, and viewport-alignment work for each person, allowing her to stay focused on evaluating and selecting participants for the panel.
\section{Cross-Page Coordination Patterns}
\label{sec:coordination-model}
\begin{figure*}[]
    \centering
    \includegraphics[width=0.95\linewidth]{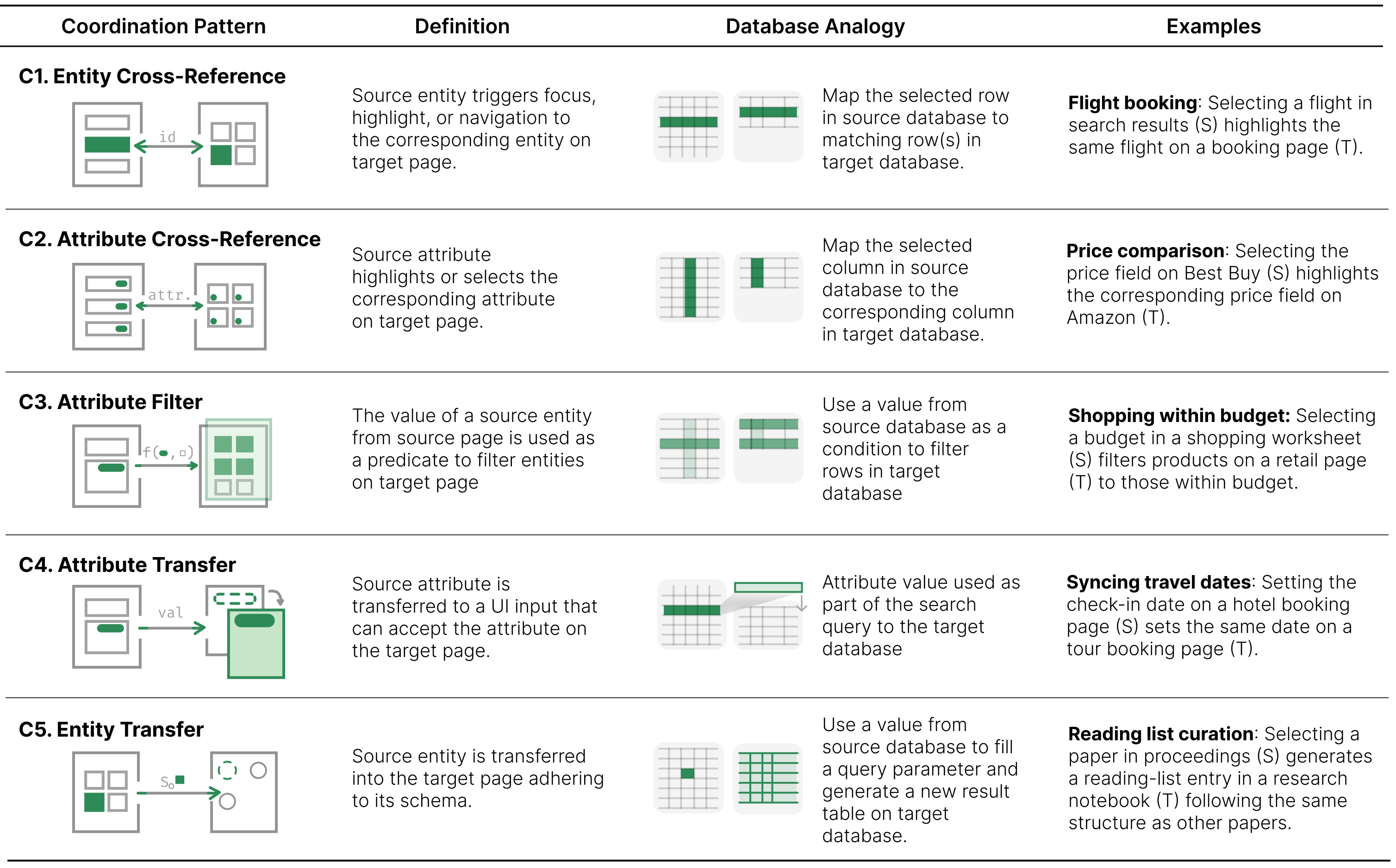}
    \caption{\textbf{Cross-Page Coordination Patterns.} Viewing web pages as databases, we categorize cross-page coordination into Entity Cross-Reference (C1), Attribute Cross-Reference (C2), Attribute Filter (C3), Attribute Transfer (C4), and Entity Transfer (C5).} 
    \label{fig:pattern-table}
    \Description[Table of five cross-page coordination patterns]{Table of five cross-page coordination patterns, with columns for pattern, definition, database analogy, and example. Diagrams use green highlights and arrows to connect source (S) and target (T) elements. C1, Entity Cross-Reference: selecting a source entity focuses, highlights, or navigates to a corresponding target entity, analogous to matching database rows; for example, highlighting the same flight across search and booking pages. C2, Attribute Cross-Reference: selecting a source attribute highlights the corresponding target attribute, analogous to matching columns; for example, highlighting price fields across Best Buy and Amazon. C3, Attribute Filter: a source value supplies a predicate for filtering target entities, analogous to filtering database rows; for example, using a worksheet budget to filter retail products. C4, Attribute Transfer: a source attribute value populates a compatible target input, analogous to supplying a search-query value; for example, transferring a check-in date from a hotel page to a tour booking page. C5, Entity Transfer: a source entity is transferred into the target page according to its schema, analogous to supplying a query parameter to generate a result table; for example, adding a paper from conference proceedings to a research notebook's reading list.}
\end{figure*}

% \rima{to connect with this section better, i would move the current 2.1 to be the 2.5 }

% Users' behaviors on the web often involve operations across multiple heterogeneous pages. To define the problem space this paper aims to support, we first introduce a catalog of cross-page coordination patterns. In what follows, we first define how we model a page, then detail the five coordination patterns, each describes to a distinct correspondence between pages.

% Users' behaviors on the web often involve operations across multiple heterogeneous pages. To characterize this space, we introduce a catalog of cross-page coordination patterns. We first define how we model a page, then present five coordination patterns, each capturing a distinct type of correspondence across pages.
Supporting cross-page coordination requires identifying how information on one page relates to information and actions on another. We use these correspondences as a basis for a catalog of five coordination patterns, each mapping a configuration of page models and their correspondences to applicable coordination actions. We first define the page model, then describe each pattern.

% We initially sought to derive coordination patterns from existing websites. However, cross-page coordination is not yet a systematic design paradigm, which makes the available corpus limited. Consequently, patterns observed in the wild would likely reflect architectural contingencies rather than a generalizable underlying coordination primitives.
% Instead, we draw inspiration from coordinated multiple views (CMV) research, where visualizations are coordinated by propagating structured data elements—such as selected records or joined subsets—across views~\cite{snapTogether2000}.  

% We extend that insight beyond data visualization to web pages: if pages are structured representations of data, then cross-page coordination can be derived from the fundamental structural elements of that data. Cross-page coordination and then be reduced to a small set of irreducible propagation types.

\subsection{Pages Viewed as Databases}
% We conceptualize each page as a \textit{structured representation that encodes a subset of entities and attributes under a specific schema.} 
We conceptualize each page as a \textit{structured representation that encodes the underlying data with a specific schema.} While not all web pages are driven by relational databases, most could be viewed from a database perspective, where each page contains entities $\mathcal{E}$ (i.e., rows), such as products and restaurants; and attributes $\mathcal{A}$ (i.e., columns), such as price and location. With this perspective, we can describe cross-page coordination with the following formalization:
% \heading{Source Page} a dataset $D_s \subseteq \mathcal{E}_s \times \mathcal{A}_s$ under schema $S_s$

% \heading{Target Page} a dataset $D_t \subseteq \mathcal{E}_t \times \mathcal{A}_t$ under schema $S_t$

% \heading{Cross-page Coordination} an operation that maps a user's expressed interest on the source to entities on the target page.

\begin{itemize}[leftmargin=*, itemsep=2pt, topsep=3pt]
    \item \textbf{Source Page:} a dataset $D_s \subseteq \mathcal{E}_s \times \mathcal{A}_s$ under schema $S_s$
    \item \textbf{Target Page:} a dataset $D_t \subseteq \mathcal{E}_t \times \mathcal{A}_t$ under schema $S_t$
    \item \textbf{Cross-page Coordination:} an operation mapping a user's expressed interest on the source page to the target page.
\end{itemize}

\subsection{Coordinations as Database Operations}

% Under the relational abstraction described above, cross-page coordination reduces to: \textit{What relational construct can be mapped from Page A to Page B?} This leads to five coordination patterns.

% Under the relational abstraction described above, cross-page coordination reduces to: \textit{What relational construct can be mapped from Page A to Page B?} This leads to five coordination patterns.

Taking this database-oriented view, we can comprehensively model cross-page coordination by examining the possible operations across two databases. This leads to five primitive coordination patterns. More complex coordination behaviors can be understood as compositions of these operations. In practice, a single task often involves multiple patterns simultaneously or sequentially.
%, which we demonstrate in Section \ref{sec:blah}.

\heading{C1. Entity Cross-Reference ($ \mathcal{E}_s \leftrightarrow \mathcal{E}_t$)} 
The user’s interest is at the entity level, analogous to selecting a row in a database table. Coordination establishes a correspondence between entities in $\mathcal{D}_s$ and $\mathcal{D}_t$ via primary-key-based joins. This includes one-to-one mappings (e.g., selecting a specific flight in a search result highlights the same flight in a booking tab) and one-to-many relationships (e.g., selecting a faculty member in a directory navigates to a list of all their publications in a proceedings). At the interface level, this appears as selecting an entity in one page to focus, highlight, or navigate to related instances in another.

\heading{C2. Attribute Cross-Reference ($ \mathcal{A}_s \leftrightarrow \mathcal{A}_t$)} In this case, the user’s interest is not in a particular entity, but in a specific attribute dimension, analogous to selecting a particular property in a dataset. Selecting an attribute in $D_s$ establishes semantic alignment with a corresponding attribute in $D_t$. In many cases, attribute names may differ across schemas (e.g., \texttt{price} and \texttt{cost}), coordination operates at the level of shared meaning rather than literal key matching. At the interface level, this may appear as highlighting an attribute (e.g., \texttt{price} or \texttt{rating}).

\heading{C3. Attribute Filter ($ \mathcal{A}_s \rightarrow \mathcal{E}_t$)} This pattern corresponds to a selection operation parameterized by a value from the source page. Formally, a value from an attribute in $D_s$ is interpreted as a predicate over $D_t$, selecting a subset of the target dataset. Unlike C2, C3 uses the source value to filter $D_t$ rather than identify corresponding target attributes.

\heading{C4. Attribute Transfer ($ Q_t: \mathcal{A}_s \rightarrow \mathcal{D}'_t$)} Interfaces often expose \textit{query-construction fields} that dynamically generate datasets (e.g., search bars, date pickers). In such cases, the target page defines a parameterized query template whose results are materialized upon instantiation. A value derived from the source page may populate such a field, thereby generating a new dataset. For example, selecting a faculty member's name may automatically populate a search field that retrieves publications from Google Scholar.

\heading{C5. Entity Transfer ($S_t: \mathcal{E}_s \rightarrow \mathcal{E}'_t$)} Entities from $D_s$ is transformed and inserted into a different schema $S_t$. Unlike direct correspondence (C1), the selected entity from  $D_s$ does not already exist in $D_t$, thus requires re-encoded under the target schema. At the interface level, this occurs when selected entities are extracted and reformatted into a new structural representation such as reformatting paper metadata into Bib\TeX{} format and adding deadlines from a course syllabus to a calendar.

These five patterns arise from a structural analysis of cross-page coordination under a relational abstraction, capturing user interest at the level of entities (C1, C5), attributes (C2, C3), and instantiated values (C3, C4). 
% In \cref{sec:tech-eval}, we evaluate how well an end-to-end pipeline can surface coordination opportunities corresponding to these patterns.

% , and in Section \ref{sec:discussion} we discuss limitations and possible extensions.
\section{Technical Pipeline}
\label{sec:technical-pipeline}
\begin{figure*}[t]
    \centering
    \includegraphics[width=0.9\linewidth]{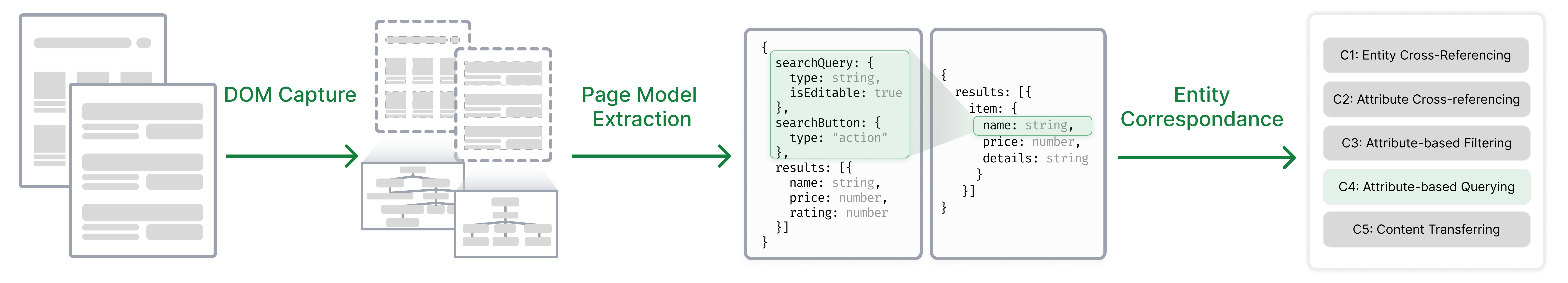}
    \caption{\textbf{\sys{}'s Technical Pipeline Overview.} For each page, the pipeline (1) captures its DOM, and (2) extracts its page model. It then uses page model correspondences to surface coordination opportunities against the five coordination patterns.}
    \label{fig:tech-pipeline}
    \Description[Left-to-right diagram of Cling's technical pipeline]{Left-to-right diagram of Cling's technical pipeline. The system captures the DOMs of two webpages and extracts structured page models containing data types and interaction properties. Highlighted fields illustrate a correspondence between an editable search input on one page and an item's name on another. These correspondences are used to identify opportunities across five coordination patterns, with attribute-based querying highlighted.}
\end{figure*}

Building on the coordination patterns, we implement a technical pipeline to construct schema-based page representations and derives coordination opportunities from them, as illustrated in \cref{fig:tech-pipeline}.

For each page, the system captures and cleans its DOM to construct a page model comprising entities and their relationships.
HTML structures varies substantially independently developer sites: generic tags such as \texttt{<div>} are often repurposed for arbitrary interface semantics, accessibility annotations are inconsistent, and markup conventions differ widely across organizations.
This heterogeneity makes it brittle to construct page models with rule-based approaches.
We therefore use LLM to extract coordination-salient semantics from the underlying markup, transforming heterogenous webpages into consistent page models.

The system then identifies correspondences across page models and maps them to opportunities for surfacing the coordination patterns outlined in \cref{sec:coordination-model}.
We adopt a heuristic-based approach, which is beneficial for real-time, interactive use.  
Specifically, the system uses a set of lightweight signals to infer cross-page correspondences to surface coordination opportunities.

\subsection{Per-Page Representation Construction}
We represent each page as a set of \emph{entities}: units of information that users interpret and operate on.

\subsubsection{Entity model}
Each entity consists of a \emph{schema}, a runtime \emph{value}, and \emph{reference} to the page's DOM.

\noindent\textbf{Schema:} 
Each entity's schema specifies its data structure and coordination-relevant interaction properties.

\begin{itemize}[
    label=\textendash,
    leftmargin=1.5em,
    topsep=3pt,
    itemsep=3pt,
    parsep=0pt
]

\item \textit{Data.} 
The schema specifies the entity's \textit{type}.
Types may be primitive (e.g., \texttt{string}, \texttt{date}) or composite (e.g., object, list); we refer to named fields of a composite entity as \textit{attributes} (e.g., price within a product entity).

\item \textit{Interaction.} 
The schema encodes properties and roles that determine how an entity can participate in coordination.
\textit{isEditable} and \textit{isExtendable} indicate whether values can be modified or added through the interface, \textit{isIdentifier} marks attributes that uniquely identify their enclosing entities, and \textit{role} distinguishes data-bearing entities from action entities (e.g., submit buttons).

\end{itemize}

\noindent\textbf{Runtime Value:} The runtime value contains the data currently associated with the entity.
For lists, the system infers one shared schema and instantiates it with the values extracted from each list item.

\noindent\textbf{Reference:}
Each entity retains one or more handles to the interface objects through
which it is presented or manipulated.
For webpages, these handles are references to DOM elements.
Reference allows the system to map user interactions on the live page to entities and to execute coordinations back to their corresponding interface objects.

\subsubsection{Page Model Extraction}
Once a page loads, the system scans its live DOM and CSS selectors to identify salient regions (e.g., lists of similarly-structured elements, standalone elements with many nested child elements, forms, and submit buttons), while removing irrelevant elements~(e.g., non-content \texttt{<script>} and decorative \texttt{style} tags).

It characterizes each salient regions and selects representative items from each list based on how common and structurally rich they are (i.e., a typical item with the most common structure and examples of other variants that contain substantial text, fields, or links).
An LLM reasons over such regions to infer the page's purpose and determine whether each region has potential to participate in cross-page coordination.

For each retained list, the LLM infers a shared item schema that captures its common attributes and meaningful structural variations.
For each list of elements, the LLM derives a schema that unionizes common attributes with optional attributes that are only seen over some list elements.
The system then maps each schema attribute to the corresponding DOM element within every list item.
It uses these mappings to extract the item's runtime values and retains references to the matched DOM elements.
Standalone entities, forms, and actions are interpreted separately and incorporated into the same page model with their runtime values and DOM references.

Page-model extraction is performed independently for each page, allowing pages to be added to or removed from a cross-page coordination without reprocessing other pages.
Completed page models are cached by URL.
For paginated pages that share the same list structure, the system can reuse the inferred list schema while extracting the runtime values and DOM references of each page anew.

\subsection{Identify Coordination Opportunities}
\label{sec:corresponsence}
With page models of all participating webpages, the system looks for each coordination pattern we formalized~(\cref{sec:coordination-model}). We outline the heuristics for recognizing these patterns below.

% \heading{C1. Entity Cross-Reference}
% \begin{enumerate}[
%     label=\textbf{C\arabic*.},
%     align=left,
%     leftmargin=2em,
%     labelsep=-1em,
%     itemsep=8pt,
%     topsep=4pt,
%     parsep=0pt,
%     itemindent=!,
% ]
\begin{enumerate}[
    label=\textbf{C\arabic*.},
    align=left,
    leftmargin=0pt,
    labelindent=0pt,
    labelwidth=1.5em,
    labelsep=0.4em,
    itemindent=!,
    listparindent=0pt,
    itemsep=8pt,
    topsep=4pt,
    parsep=0pt
]
\item \textbf{Entity Cross-Reference.}
C1 identifies entities on the target page that refer to the same underlying entity. 
\begin{list}{}{%
    \setlength{\leftmargin}{1em}
    \setlength{\rightmargin}{0pt}
    \setlength{\topsep}{3pt}
    \setlength{\itemsep}{0pt}
    \setlength{\parsep}{0pt}
    \setlength{\partopsep}{0pt}
}
\item
$H_1(e_s, \mathcal{D}_t)$ returns entities $e_t \in \mathcal{D}_t$ satisfying both conditions:
    (1) $e_s$ and $e_t$ have the same type or semantically adjacent attribute names;
    (2) They share at least one matched attribute pair $(a_s, a_t)$
    with similar values, where at least one of the two attributes
    is an identifier.
\end{list}

% \begin{list}{}{\leftmargin=1.8em \rightmargin=1.8em    \itemsep=0pt
%     \topsep=0pt
%     \parsep=0pt
%     \partopsep=0pt}
% \item
% ${H_1}(e_s, \mathcal{D}_t)$: return entities $e_t \in \mathcal{D}_t$ s.t.
% $e_s$ and $e_t$ are of the same type or have similar attribute names, and share at
% least one matched attribute pair $(a_s, a_t)$ with similar values where at least one
% attribute is an identifier.
% \end{list}

% \heading{C2. Attribute Cross-Reference}
\item \textbf{Attribute Cross-Reference.}
C2 maps a source attribute to corresponding attributes across on the target page. 

\begin{list}{}{%
    \setlength{\leftmargin}{1em}
    \setlength{\rightmargin}{0pt}
    \setlength{\topsep}{3pt}
    \setlength{\itemsep}{0pt}
    \setlength{\parsep}{0pt}
    \setlength{\partopsep}{0pt}
}
\item
$H_2(a_s, \mathcal{D}_t)$ returns target
attributes $a_t$ in $\mathcal{D}_t$ satisfying both conditions:
    (1) $a_s$ and $a_t$ have semantically adjacent attribute names;
    (2) Each attribute occurs across multiple entities in their respective schemas.
\end{list}
% \end{enumerate}
% ${H_2}(a_s, \mathcal{D}_t)$ (Attribute Matching) returns attributes $a_t \in \mathcal{D}_t$ s.t.
% $a_s$ and $a_t$ have similar attribute names and both recur across multiple entities in
% their respective schemas.
% \end{list}

% For example, when the source is an \texttt{address} attribute of an ice
% cream shop, the system identifies the \texttt{address} attribute for all restaurants on the target page.

% \heading{C3. Attribute Filter}
\item \textbf{Attribute Filter.}
C3 uses a source attribute as a constraint for filtering entities on another page. 
% For example, a source \texttt{address} attribute can filter results on the target page to those within some distance radius from the source address. 
% It applies a type-specific comparison, such as date proximity or geographic distance, to filter or rank target entities.
Let ${S_t}(v_s, v_t)$ denote a type-specific comparison function that scores the
relationship between a source value $v_s$ and a target value $v_t$. C3 uses a configurable threshold
$\tau_t$ to convert this score into a filter predicate.
% , and may additionally use the
% score itself to rank target entities.

\begin{list}{}{\leftmargin=1.8em \rightmargin=1.8em\itemsep=0pt
    \topsep=0pt
    \parsep=0pt
    \partopsep=0pt}
\item
% ${H_3}(a_s, \mathcal{D}_t)$: let $M \leftarrow H_2(a_s, \mathcal{D}_t)$; return entities
% $e_t \in \mathcal{D}_s$ s.t. $e_t$ contains some $a_t \in M$ with
% $S_t(\mathrm{val}(a_s), \mathrm{val}(a_t)) \ge \tau_t$.

$H_3(a_s, \mathcal{D}_t)$ first identifies matching attributes $M = H_2(a_s, \mathcal{D}_t)$, then returns entities $e_t \in \mathcal{D}_t$ satisfying both conditions:
(1) $e_t$ contains an attribute $a_t \in M$;
(2) For that attribute, $S_t(\mathrm{val}(a_s), \mathrm{val}(a_t)) \ge \tau_t$, where $\mathrm{val}(a)$ denotes the value of attribute $a$.
\end{list}

% \heading{C4. Attribute Transfer}
\item \textbf{Attribute Transfer.}
% C4 transfers an attribute's value from one page into a compatible editable entity on another page.
% For example, selecting ``red'' on one page may populate a ``color'' field on another.
C4 transfers a source attribute's value into a compatible editable field on the target page.

% \begin{list}{}{\leftmargin=1.8em \rightmargin=1.8em\itemsep=0pt
%     \topsep=0pt
%     \parsep=0pt
%     \partopsep=0pt}
% \item
% ${H_4}(a_s, Q_t)$: return editable fields $q_t \in Q_t$ s.t. 
% $a_s$ and $q_t$ are either type-compatible or have similar attribute names.
% \end{list}
\begin{list}{}{%
    \setlength{\leftmargin}{1em}
    \setlength{\rightmargin}{0pt}
    \setlength{\topsep}{3pt}
    \setlength{\itemsep}{0pt}
    \setlength{\parsep}{0pt}
    \setlength{\partopsep}{0pt}
}
\item
$H_4(a_s, Q_t)$  returns editable
fields $q_t \in Q_t$ satisfying at least one condition:
    (1) $a_s$ and $q_t$ are type-compatible;
    (2) $a_s$ and $q_t$ have semantically adjacent attribute names.
\end{list}

% \heading{C5. Entity Transfer}
\item \textbf{Entity Transfer.}
% C5 transfers a source entity into an extendable target entity, such as a spreadsheet, calendar, or map. Unlike C1, it transforms the source entity to fit the target structure rather than matching it to an existing target entity.
C5 transfers a source entity into an extendable target entity, such as a spreadsheet, calendar, or map. Unlike C1, it adapts the source to the target structure rather than matching an existing target entity.

\begin{list}{}{%
    \setlength{\leftmargin}{1em}
    \setlength{\rightmargin}{0pt}
    \setlength{\topsep}{3pt}
    \setlength{\itemsep}{0pt}
    \setlength{\parsep}{0pt}
    \setlength{\partopsep}{0pt}
}
\item
$H_5(e_s, S_t)$ returns extendable
target structures $s_t \in S_t$ for which $e_s$ can be transformed
into a format compatible with the target schema.
\end{list}

% \begin{list}{}{\leftmargin=1.8em \rightmargin=1.8em\itemsep=0pt
%     \topsep=0pt
%     \parsep=0pt
%     \partopsep=0pt}
% \item
% ${H_5}(e_s, S_t)$: return extendable target structures $s_t \in S_t$ s.t. $e_s$ can
% be transformed into a format compatible with $s_t$.
% \end{list}

$H_5$ surfaces target structures that can accept a transformed version of the selected source entity $e_s$. When invoked, the system supplies source and target context---including the selected entity, its parent, and the surrounding target entities---to transform $e_s$ into a representation compatible with $s_t$.

\end{enumerate}

% $H_5$ surfaces target entities that can accept a transformed version of the selected
% source entity $e_s$. When $H_5(e_s, S_t) \neq \emptyset$ and the user invokes $H_5$, the system provides the selected source entity $e_s$ and its parent entity as source context, and the target extendable entity and its surrounding entities as target context, so that the source can be transformed into a representation compatible with $S_t$.

\subsection{Coordination Generalization}
\label{sec:binding-generalization}
Coordination opportunities are first identified between a source-target entity pair. The system then generalizes the pairwise coordination to analogous entities in the source and target schemas via structural relationships such as array membership, entity containment, and relative paths within composite entities, enabling reuse of the established coordination for other suitable pairs.

\subsection{Incremental Entity Updates}
After constructing a page model, the system keeps it synchronized with the live page by monitoring the DOM elements that ground its entities.
When the DOM changes, these references identify the affected portion of the page model, allowing the system to update it without re-extracting the page.

For an updated list item, the system reapplies the list's existing schema to the affected DOM subtree.
The item's \textit{identifier} attribute determines whether the result updates an existing entity or creates a new one, and removing the subtree removes the corresponding entity.
For an updated object attribute, the system parses the updated content using its existing type and updates only that attribute.
If its referent is replaced, the system reconnects it to the replacement element.
If the updated content cannot be interpreted as the attribute's existing type, parses it as a new entity and removes the old entity.

\subsection{Implementation}
The pipeline is implemented in TypeScript as part of an Electron application.
The current implementation uses \texttt{gemini-3.5-flash-lite} by default.
Semantic comparisons between entities are also performed locally using \texttt{all\allowbreak-MiniLM-L6\allowbreak-v2} with cosine similarity.

\section{\sys{}: User Interfaces and Interactions}
\label{sec:cling}
\begin{figure*}[t]
    \centering
    \includegraphics[width=0.96\linewidth]{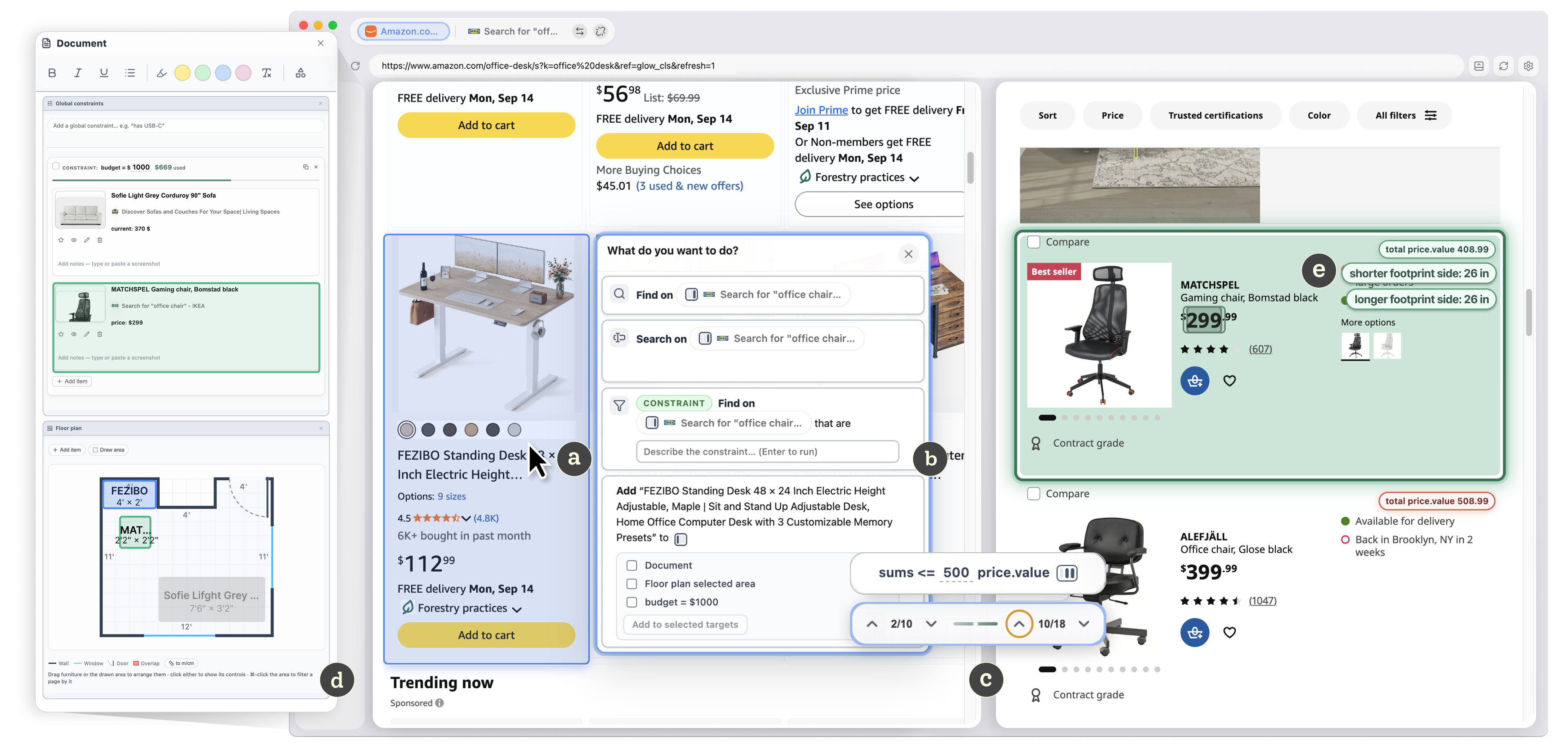}
    \vspace{-10px}
    \caption{\textbf{Using \sys{}.} The user selects an desk as a source by \texttt{cmd+clicking} it (a). \sys{} surfaces coordination opportunities in a pop-up panel, where the user can select actions and specify their criteria (b). Once accepted, active coordination actions appear in a ``binder'' between the participating pages (c). To extend coordination across additional pages and information sources, \sys{} provides a side panel for saving and reusing actions (d) and retrieves external attributes to supplement the page model (e).}
    \label{fig:flow}
      \Description[The demonstration of using Cling system for cross-page coordination task]{The demonstration of using Cling system for cross-page coordination task. The screenshots show the workflow of using Cling to coordinate furniture shopping across an Amazon desk listing and an IKEA chair catalog displayed side by side. On the left side is the Amazon desk listing, the user's mouse is hovering on desk item, highlighting in blue; on the right side is the IKEA char catalog,  a matching chair under budget constraint is highlighted in green. Five callouts illustrate the interface: (a) a desk selected as the source is outlined in blue; (b) a pop-up panel offers actions to find, search, specify constraints, or add the item to a document; (c) a binder between the pages displays an active combined-price constraint of at most 500 dollars and navigation controls; (d) a side panel contains saved furniture items, a shared budget, and a floor plan; and (e) attribute badges on chair listings show combined prices and retrieved footprint dimensions.}
\end{figure*}

We designed \sys{} to reify cross-page coordination by making implicit relationships between heterogeneous pages explicit. \cref{fig:flow} illustrates the end-to-end flow from intent specification to how coordination opportunities are surfaced and executed.

\subsection{Specifying Coordination Intent and Surfacing Coordination Opportunities}
\sys{} allows users to place two pages into side-by-side views to indicate interest in coordinating between them. 
For any pair of pages with extracted page models, \sys{} does not speculatively correspond compatible entities to compute all coordination opportunities because it would warrant exhaustively exploring a large correspondence space.
Instead, it waits for users to specify their coordination intent by \texttt{cmd+clicking} a DOM element on the page. \sys{} uses the bidirectional reference to the page model to identify its corresponding entity as the source entity. 
Then, \sys{} runs the correspondence heuristics to identify all possible coordination opportunities involving the specified source entity.

Once coordination opportunities are surfaced following a user specification, \sys{} presents them in a popover menu next to the selected DOM element (\cref{fig:flow}a), for users to select the intended coordination. 
When proposed, each coordination pattern in \cref{fig:pattern-table} deterministically maps to one option on the popover menu.
Some coordination patterns allow the user to further refine the coordination requirement on the popover.
For example, when attribute filter (C3) is proposed  (\cref{fig:flow}b), the user can specify the constraints with more concrete criterion (e.g., ``total price below 500") in natural language.
% Coordination pattern in \cref{fig:pattern-table} is represented by a coordination chip, as shown in \cref{fig:flow}.

% \begin{figure}[t]
%     \centering
%     \includegraphics[width=0.98\linewidth]{assets/flow.png}
%     \caption{Flow Placeholder}
%     \label{fig:flow}
% \end{figure}

% A chip becomes colorized when its coordination pattern is available for the selected source entity; otherwise, it remains grayed out and disabled. Because attribute filter (C3) builds on attribute cross-reference (C2), where to filter, one must first select all attributes on the target page, they share a single chip. 

\subsection{Executing and Updating Coordination}
When users accepts a coordination opportunity, \sys{} parses the runtime value for the source entity and the selected coordination pattern into a script that queries the target page's page model.
Running this script over the target page model executes the selected coordination, and the the bidirectional references encoded by the page model is used to present coordination progress and results in-situ.

\sys{} presents the currently active coordination as a ``binder'' between the pair of participating pages  (\cref{fig:flow}c).
The binder displays a compact chip displaying the coordination pattern, and optionally the current filtering constraints whose numeric values are editable.
The binder also supports navigation among all current source and target entities inferred through binding generalization~\cref{sec:binding-generalization}. 
Stepping to a different target entity auto-scrolls it into view.
Stepping to a different target or updating the numeric value of the constraint updates the script used to query the target page's page model and reruns over it to updates the latest satisfying target entities.

For coordination patterns that modify the target page (e.g., changing values, inserting content, or navigating to another page), \sys{} uses a two-step execution flow. It first shows an overlay next to the target entity's corresponding DOM element to preview the effect of the change, and ask the user to confirm before committing the action. When the user confirms the same type of coordination for the same source entity once, subsequent confirmations can be optionally skipped.
For coordination patterns that only update the target page's presentation (e.g., highlighting corresponding entities or scrolling them into view), \sys{} applies the results immediately.

\subsection{Extending the Scope of Coordination}
\label{sec:extend-scope}

\sys{} extends coordination support through action reuse across additional pages and retrieval of attributes beyond the initially extracted page models.

\subsubsection{Reusing Actions Across Pages}
\label{sec:beyond-pair}
Cross-page coordination in web tasks often involves more than two pages.
To extend coordination beyond two pages arranged side by side, \sys{} allows users to save pairwise coordination actions and reuse them across additional pages.
Saved actions, such as those applying a shared budget or spatial constraints from a floor plan, appear in a side panel (\cref{fig:flow}d), where users can selectively apply them to new page pairs.

This provides a visual way to reuse actions across pages and incrementally apply multiple constraints. For example, when selecting furniture under budget and spatial constraints, users can save actions that check prices against a shared budget and dimensions against available space in a floor plan. As they browse additional listings, they can reapply these actions to assess each option against both constraints without repeatedly establishing the same cross-page relationships.

%%%% ----- PREVIEW VERSION ------ 
% Cross-page coordination in web tasks often involve more than two participating pages.
% To scale \sys{}'s coordination support beyond pairwise pages arranged side-by-side, it enables people to selectively persist pairwise coordination for reusing over additional pages.
% \sys{} presents persisted coordination instances, such as global budget or a floor plan as spatial constraints, on a side panel (\cref{fig:flow} d) for people to selectively reuse over multiple pair-wise coodinations.
% A coordination can be reused on a new page as long as the page model is compatible with the reused coordination.
% Specifically, a new page can serve as the source of a reused coordination if it contains type-compatible entities that can serve as the source entity.
% Likewise, a new page can serve as the target of a reused coordination as long as its page model contains eligible target for the coordination's underlying script to be evaluated against.

% \subsection{Scaling Coordination with External Attributes}

\subsubsection{Retrieving External Attributes}
\label{sec:beyond-attr}

Coordination actions may depend on attributes that are not available on the existing extraction of the page model. For example, when selecting furniture, 
the dimensions the user looks for may be nested in the product detail pages. Likewise, when filtering a houseplant catalog, a user may specify a ``cat-safe'' criterion that requires searching for general knowledge beyond the catalog. 
To retrieve these missing attributes, \sys{} uses the Gemini API with Google Search grounding~\cite{gemini_search_grounding} to obtain the requested attribute values for relevant entities on the page (\cref{fig:flow}e). 
% This allows the page model to accommodate users' idiosyncratic needs as they arise, extending beyond the schema and runtime values initially extracted from the current webpage.
% % A cross-page coordination
% The requirements for a coordination may involve attributes external to any participating pages.
% For example, when filtering a house plants catalog, one may specify an ``cat-safe'' criterion that is not immediately present on the page.
% \sys{} uses the \todo{google web-search api, double check the exact api name} to extract values for this attribute for each relevant entity on the page.

\subsection{Example Coordination Flows with \sys{}}

We illustrate the user-facing interaction flow of \sys{} through four examples, covering all coordination patterns (\cref{sec:coordination-model}).

\heading{Cross-referencing schools across university rankings (C1)}
Jackie puts a university ranking page and a page listing top universities for statistics major side by side. She cmd-clicks the name of a school on the general ranking page, and \sys{} highlights the C1 coordination chip in the binder and shows the school was found on the major ranking. Jackie then clicks \textit{Go To} on the chip, and \sys{} scrolls the matching school on the other page into view so she can inspect its details. As she works through schools she is considering applying to, she uses the navigation widget to move through the ranking list, skipping entries for which \sys{} reports \textit{No Match}.

\heading{Filtering restaurants by proximity to a hotel (C2/C3)}
Lily places a hotel listing page beside a restaurant listing page. She cmd-clicks a hotel's \texttt{address}, and \sys{} activates the C2/C3 coordination chip on the binder, highlighting corresponding \texttt{address} values on the restaurant page. Hovering over the chip reveals filters for location-typed attributes. She selects a radius filter and sets it to 1 mile. The chip then displays ``Within 1 mile,'' and only restaurant addresses within 1 mile of the selected hotel remain highlighted. Lily expands the navigation widget into a list showing each filtered restaurant's distance from the hotel and clicks the closest one to scroll it into view. When she moves to the next hotel, \sys{} automatically reapplies the same filter relative to the new hotel.

\heading{Querying winery availability from forecasted dates (C4)}
Nina places a weather page beside a winery booking page. She browses the forecast for a sunny Saturday and cmd-clicks a date. \sys{} activates the C4 coordination chip in the binder, indicating that the selected date can transfer to a compatible date picker on the booking page. Invoking the chip opens an overlay next to the date picker that previews the value to be inserted. After Nina confirms, \sys{} populates the field and updates the booking page to show available tasting slots for that day. When she selects another date, \sys{} reuses the same confirmation preference and directly updates the date picker to show availability for the new date.

\heading{Curating Reading List During Literature Review (C5)}
Russell places a conference proceedings page beside his research notebook, a markdown editor. As he goes through the proceedings list, he \texttt{cmd-clicks} a relevant paper. \sys{} highlights the C5 coordination chip in the binder, indicating that the selected paper can be transformed into a structure compatible with the notebook, and shows an overlay previewing the insertion. After he confirms, \sys{} inserts a new entry with the same structure, including the title, venue, and abstract. Russell then continues through the proceedings, transferring additional relevant papers the same way.
\section{Technical Evaluation}
\label{sec:tech-eval}
We conducted a technical evaluation of \sys{}'s end-to-end pipeline by first extracting page schemas and then applying schema correspondence to surface cross-page coordination opportunities. Although the evaluation focuses on surfaced opportunities, it also reflects schema extraction quality, as downstream coordination depends on it. Our evaluation was guided by the following research questions:
\mybullet{RQ1} How \textit{accurate} are the coordination opportunities surfaced by the pipeline?
\mybullet{RQ2} How \textit{useful} are the surfaced coordination opportunities in the context of the pages being coordinated?

\subsection{Methodology}
We constructed an evaluation dataset spanning three task domains commonly used in prior evaluations of web-based systems: \emph{travel planning}~\cite{paul2009cosense, morris2010wesearch, chang2019searchlens, huang2025irchiver}, \emph{shopping}~\cite{chang2020mesh, kuznetsov2022fuse, liu2022wigglite}, and \emph{information work}~\cite{kuznetsov2022fuse, shaikh2025creating, chen2023miwa, liu2022crystalline, liu2019unakite}. Within each domain, we identified five representative tasks. For each task, we selected three web pages that were central to the completion of the task and that exhibited diverse interface structures. We then combined these pages into pairs, yielding three pairs per task. In total, this produced 45 page pairs. To ensure stable evaluation conditions, we archived each page by downloading its HTML and associated assets and serving them locally, thereby avoiding nondeterministic changes in live web content. We then ran the schema extraction pipeline for each archived page using gemini-2.5-flash-lite~\cite{comanici2025gemini}.

For each page pair, we sampled representative source entities from the extracted schemas by selecting one instance of each distinct entity type on the page. For each sampled source entity, we ran the pipeline to surface all possible coordination opportunities with the paired page. When appropriate, we manually added reasonable task-level constraints, such as a shared budget across two shopping pages, to reflect realistic coordination scenarios. This process yielded 220 coordination opportunities. Across the 5 patterns, the dataset includeds 17 Entity Cross-reference (C1), 87 Attribute Cross-reference (C2), 30 Attribute Filter (C3), 71  Attribute Transfer (C4), and 15 Entity Transfer (C5). The large numbers of C2 and C4, compared to the others, are expected, as entities often contain multiple attributes. Two external human coders independently annotated each surfaced coordination opportunity for \textbf{accuracy}, whether the pipeline correctly identified the opportunity, and \textbf{usefulness}, how helpful the opportunity would be in the context of the paired pages and task scenario. Usefulness was rated on a 5-point ordinal scale.

Our accuracy measure evaluates the correctness of surfaced opportunities, rather than recall over all potentially useful coordinations.
Measuring recall would require a ground truth of all salient coordination opportunities across all possible tasks involving a page pair, which is difficult to define because relevance depends on users’ intents, constraints, and workflows.

To assess whether the technical pipeline supports interactive use during browsing, we measured entity model extraction time for each of the 48 webpages in the evaluation suite over three independent runs.

\subsection{Results}
For accuracy, the two coders achieved high inter-rater reliability across the 220 evaluated coordination opportunities. We report Gwet's AC1~\cite{gwet2008computing}, as very high raw agreement in several categories can make measures such as Cohen's $\kappa$ unstable~\cite{cohen1960coefficient}. Overall, coders achieved Gwet's AC1 = 0.910 with 92.3\% observed agreement, indicating substantial agreement.
We did not seek consensus on usefulness because, unlike correctness, it was intentionally rated with respect to each coder's own lived experience.
% Therefore, we report the distribution of individual ratings rather than treating disagreement as annotation error.

\heading{Accuracy of Surfaced Opportunities}
Overall, the pipeline surfaced coordination opportunities with high accuracy, achieving an average accuracy of \textbf{92.5\%} across all 220 opportunities. Broken down by coordination pattern type, the average accuracies were 94.1\%, 92.5\%, 85.0\%, 94.4\%, and 96.7\% for C1 through C5, respectively.

\heading{Usefulness of Surfaced Opportunities}
Surfaced opportunities were generally rated as useful, with a mean score of \textbf{4.48}/5. Mean ratings for C1--C5 were 4.94, 4.08, 4.57, 4.72, and 4.93, respectively.

\heading{Entity Model Extraction Latency}
Across individual pages, the mean extraction time was 5.82 seconds (95\% CI: 5.31–6.33 seconds), with a median of 5.02 seconds and a 95th percentile of 11.46 seconds.

\subsection{Takeaways}
The results suggest that the pipeline can reliably surface coordination opportunities that are both accurate and useful across coordination patterns. Accuracy was high overall and strongest for C1 and C5, indicating that the pipeline performs especially well when coordinating at the entity-level. Usefulness was also high overall, but varied more across coders and coordination patterns, suggesting that perceived value depends more on task context and prior habits than correctness does.

Most inaccuracies stemmed from attribute misalignment. Schema extraction sometimes retained under-specified labels such as \texttt{name}, even when multiple entity types on the same page (e.g., foods and restaurants) exposed that label, causing  incorrect matches. Repeated entities on a page also did not always share identical attribute structures: some items included additional or variant attributes, such as \texttt{discounted to} alongside \texttt{price}. This could cause the pipeline to miss the intended correspondence or coordinate on an available but less relevant attribute (e.g., using the listed rather than discounted price when applying a budget constraint during shopping). C3 was further affected by our use of a single semantic similarity threshold, which could be either too strict (e.g., failing to match product titles when one page omitted a brand name) or too permissive (e.g., matching address strings that shared city, state, and ZIP code despite referring to different locations).

Variation in usefulness ratings reinforces our decision to present coordination opportunities in a lightweight, non-distracting manner rather than as a persistent or intrusive feature. This is especially important for C2, which is inexpensive to propose and thus easy to surface frequently, but may not always be useful for the user.

% \subsection{Comparative evaluation among LLM models}
% kinda optional
\section{User Study}
\label{sec:user-study}

We conducted a two-part user study of \sys{}, examining its effectiveness in structured cross-page coordination tasks and its role in broader exploratory information activities.
We sought to answer the following research questions:
\mybullet{RQ1} How effectively and efficiently does \sys{} support cross-page coordination?
\mybullet{RQ2} What are the \textit{values} and \textit{limitations} do people experience with \sys{}'s in-situ, small-step automation during cross-page tasks?
\mybullet{RQ3} How do people incorporate cross-page coordination support into broader exploratory tasks?

% \todo{write about scratchpad for all conditions for ecological validity (cite people externalize working memory with spreadsheet etc), but disable features that could automate the task in such structures.}

% \todo{why no ai baseline in part 1: also, what we established earlier in intro "e2e automation eg computer use is great at completing long-horizon coordination tasks", that has been demonstrated by the computer use agent community to truly evaluate "how e2e computer use across all coordination low level operations" require a much bigger benchmark with more diverse task domains, workloads, scale of the task, and the cross-task findings on length of prompt etc across the 3 selected task are not generalizable. this is also becoming evaluation of the model and its harness than what our paper is about.}

\subsection{Participants}
We recruited 12 participants (6 female, 6 male, aged 20–28) through internal communication channels within a large public university. Participants included undergraduate and Ph.D. students. All participants use browsers on a daily bases and find themselves using multiple pages to complete a single task many times a day. All sessions were conducted in person and lasted approximately 100 minutes. Each session was screen- and audio-recorded. Participants received a 35 USD Amazon gift card as compensation.

\subsection{Prototype Conditions}
All three conditions used the same underlying Electron/Chromium browser shell, which supported side-by-side page views and provided lightweight spreadsheet, calculator, and rich-text editor tools. The conditions differed only in the coordination and automation support layered on top of this common environment.

\paragraph{\manual{}.}
The manual condition used the base browser shell without any \sys{}-specific cross-page coordination or AI automation support.

\paragraph{\ai{}.}
The \ai{} agentic condition augmented the same browser shell with an AI assistant implemented using Browser Use v0.13.8\footnote{\url{https://pypi.org/project/browser-use/0.13.8/}}, an open-source framework for connecting LLMs to web browsers, with GPT-5.6 Sol as the underlying model.
To provide a strong browser-automation baseline, we enabled DOM-based page inspection, bounded JavaScript execution for batched edits, and a scratchpad for tracking multi-step tasks.
% \footnote{\todo{The full system prompt and agent configuration are provided in Appendix~X.}}
The agent was instructed to ground task-specific claims in evidence observed during the task rather than rely solely on pretrained knowledge.

\paragraph{\sys{}.}
The \sys{} condition augmented the same browser shell with the technical backend described in~\Cref{sec:technical-pipeline} and the cross-page coordination interactions described in~\Cref{sec:cling}.

\subsection{Procedure}

The study consisted of the following phases.

\heading{Introduction (5 mins)} Participants were introduced to study context and procedure; informed consent was obtained.

\heading{Tutorial (10 min)} 
Participants went through structured tutorial for all three prototypes.
For each tutorial, the experimenter first demonstrated the relevant interactions using a separate example task, after which participants completed a brief hands-on practice task and could ask clarification questions.
Practice materials were not reused in the measured tasks.

\heading{Part 1: Structured Cross-Page Coordination (40 min)}
Participants completed three structured cross-page coordination tasks under each of the three interface conditions: \manual{}, \ai{}, and \sys{}.
We used a within-subject design where interface and task orders were completely counterbalanced across the 12 participants. Each task had three matched content variants, which were counterbalanced across interface conditions. For each participant, task order was held constant across interface conditions.
The tasks were designed around representative cross-page coordination episodes spanning all five coordination patterns introduced in~\Cref{sec:coordination-model}, while retaining objectively specified outcomes that allowed us to measure task completion, correctness and completeness, and completion time.
In the \ai{} condition, participants were instructed to use the AI assistant as the primary means of interacting with webpages, asking the agent to carry out page actions on their behalf whenever possible. They remained free to inspect the pages themselves to understand the information or verify the agent's work.

\begin{itemize}[leftmargin=*]
    \item \textbf{Cross-Page Find (C1, C2).} Cross-reference a spreadsheet of PC cases inventory with an editorial recommendation guide and mark which cases in the spreadsheet are recommended by the guide.
    
    \item \textbf{Cross-Page Filter (C2, C3).} Cross-reference landmarks and restaurants across two city guides, identify landmark--restaurant pairs satisfying a specified distance constraint, and report the qualifying pair that best satisfies the objective selection criterion. In the \manual{} condition, participants were given access to Google Maps, and \ai{} was given Google Maps API to query distance.
    
    \item \textbf{Cross-Page Transfer (C4, C5).} For each paper on a conference proceedings page, look it up on a archived Semantic Scholar interface, identify papers satisfying a citation threshold, and transfer the qualifying papers and their metadata to a spreadsheet.
\end{itemize}

To ensure stable environments, all pages were archived and served locally and their schemas were loaded from cached extractions.
Based on pilot sessions, we scoped each task to 12--16 coordination instances (e.g., 12 landmark--restaurant pairs), balancing sufficient repetition with a reasonable chance of completion in the manual condition.

After completing all tasks using one prototype, participants completed a questionnaire with 5-point Likert-scale related to their experience of working on the tasks using the prototype.

\heading{Break (5 min)}
Participants took a 5-minute break.

\heading{Part 2 Tutorial (5 min)}
Before Part 2, participants received a brief guided tutorial on the additional capabilities used in the broader furnishing tasks. 
The experimenter first introduced the interactive floor-plan widget embedded in the task document for both \ai{} and \sys{}, which allowed participants or agent to add dimensioned representations of selected furniture and arrange them spatially.
For \ai{}, participants were informed that the assistant could additionally draw on pretrained knowledge and Web search to support the broader task.
For \sys{}, the experimenter demonstrated persisting and reusing coordination beyond page pairs~(\cref{sec:beyond-pair}).
Participants briefly practiced these features using separate example materials before beginning the measured tasks.

\heading{Open-ended Floor Planning Tasks {20 min}}
Participants completed one 10-minute furnishing task with \ai{} and one with \sys{}, with condition order and room assignment counterbalanced.
In the bedroom variant, participants were asked to select a bed frame and at least one additional major item; in the living-room variant, they were asked to select a sofa and at least one additional major item.
Beyond these requirements, participants were free to determine their final choices, evaluation criteria, ordering of actions, and which pages to visit, subject only to the constraint that their selected furniture fit within the provided floor plan.

In each condition, participants could choose whether and when to use the available support---\sys{} or the AI assistant---and could otherwise browse manually.
Participants recorded their current selections in the provided document.
They were not required to complete the furnishing plan within the 10-minute period, but were asked to work toward a plan they would be satisfied with.

\heading{Questionnaire and interview (15 min)} 
After each Part 2 trial, participants completed a brief questionnaire with 5-point Likert-scale items about their experience using the prototype.
After both trials completed, we ended with a semi-structured interview for additional qualitative feedback.

\subsection{Results and Findings}
% \subsectionnocolon{he}

% \subsubsection{example}
% \zh{have subsubsection: benefits and limitations, and have bolded headings under them}
% \subsectionnocolon{Benefits}
% \subsectionnocolon{he}
% \subsectionnovspace{hi}

\begin{figure*}[!ht]
    \centering
    \includegraphics[width=0.8\linewidth]{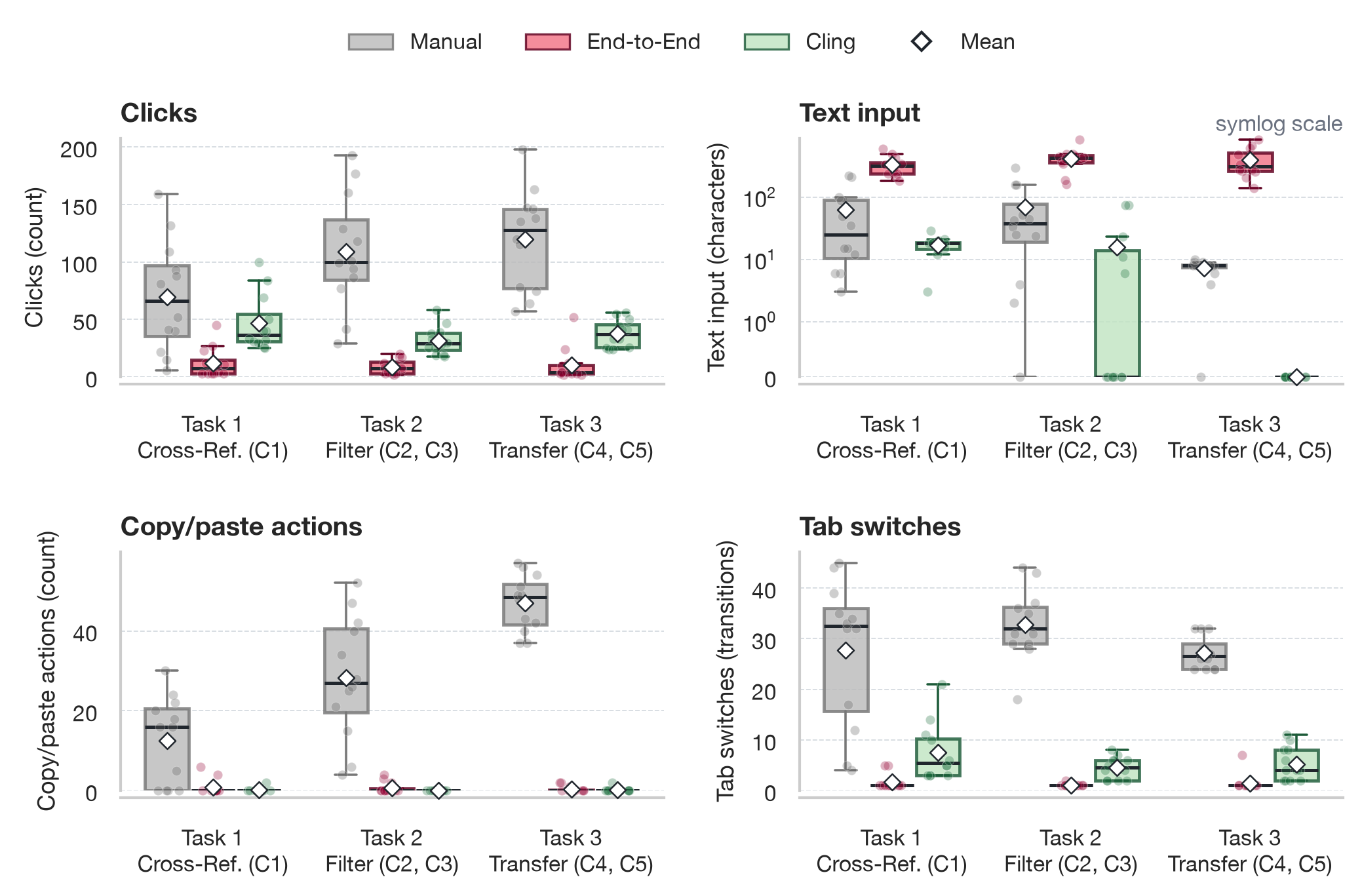}
    \caption{\textbf{Interaction Counts Across the Three Part~1 Tasks.} Compared with \manual{}, \sys{} reduced clicks, typed characters, copy/paste actions, and tab switches. \ai{} required fewer clicks and tab switches than \sys{}, but substantially more typed characters for prompting.}
    \label{fig:interactions}
    \Description[Four panels of box plots compare Manual (gray), End-to-End (pink), and Cling (green) across three tasks]{Four panels of box plots compare Manual (gray), End-to-End (pink), and Cling (green) across three tasks: cross-referencing (C1), filtering (C2, C3), and transferring (C4, C5). Panels show clicks, text input in characters, copy/paste actions, and tab switches. Dots represent individual observations, and diamonds indicate means. Text input uses a symmetric logarithmic scale. Across all tasks, Cling requires fewer clicks and tab switches than Manual, while End-to-End requires the fewest. End-to-End requires substantially more text input than either other condition. Copy/paste actions are near zero for both Cling and End-to-End, whereas Manual requires increasingly more from Task 1 to Task 3. Cling requires no text input in Task 3.}
\end{figure*}

\begin{figure}[!h]
    \centering
    \includegraphics[width=0.85\linewidth]{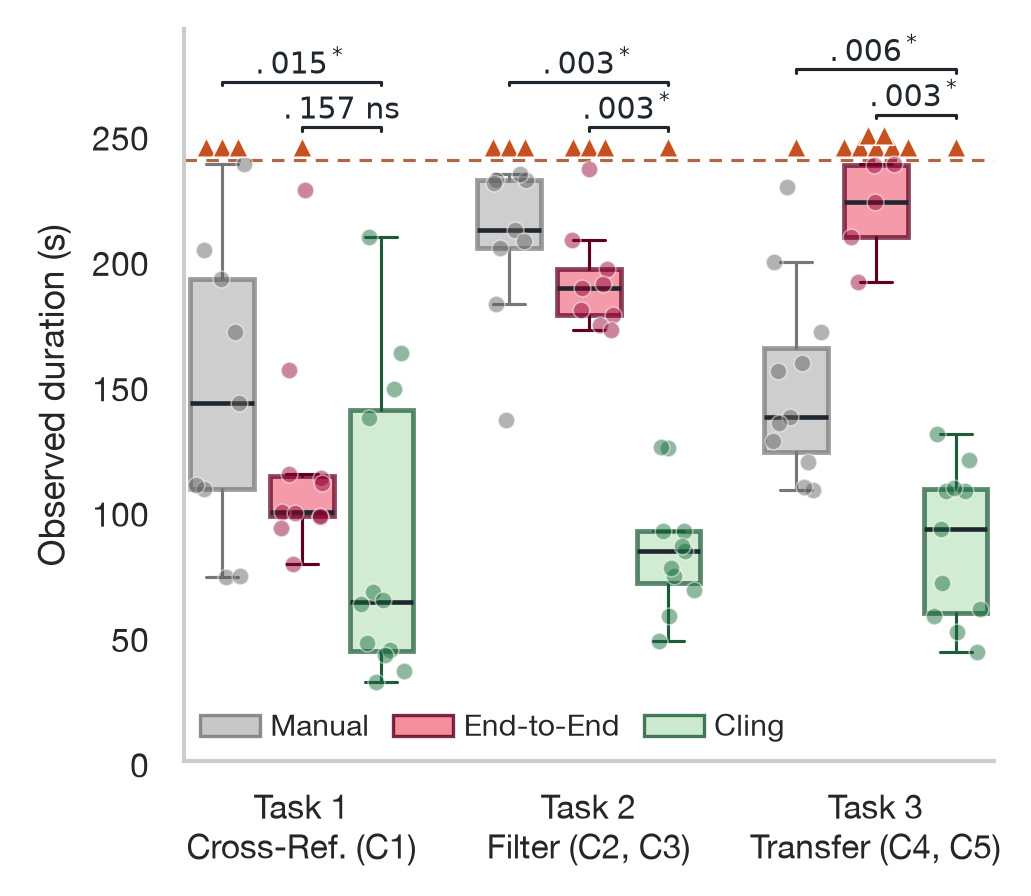}
    \caption{\textbf{Task Durations Across the Three Part~1 Tasks.} With timed-out trials assigned 240~s, \sys{} reduced overall geometric mean duration by 51.7\% relative to \manual{} and 53.2\% relative to \ai{}.}
    \label{fig:completion-time}
    \Description[Box plots compare task durations in seconds]{Box plots compare task durations in seconds for Manual (gray), End-to-End (pink), and Cling (green) across cross-referencing, filtering, and transferring tasks. Dots show individual observations; orange triangles mark timed-out trials above a dashed line at the 240-second threshold. Cling has the lowest median duration in all three tasks. Paired comparisons show significantly shorter durations for Cling than Manual in Tasks 1, 2, and 3, with adjusted p-values of .015, 0.003, and 0.006, respectively. Cling is also significantly faster than End-to-End in Tasks 2 and 3, both p = 0.003, but not in Task 1, p = 0.157.}
\end{figure}

\heading{\sys{} Offers Effective and Efficient Cross-Page Coordination Support}
\cref{fig:interactions} shows that, across the three Part~1 tasks, \sys{} substantially reduced the primary manual operations required to complete the workflows. Compared with \manual{}, \sys{} required 61.1\% fewer clicks (38.7 vs.\ 99.4 per task; paired-bootstrap 95\% CI for the reduction $=[49.5\%, 68.6\%]$), 76.6\% fewer typed characters (10.9 vs.\ 46.7; 95\% CI $=[56.1\%, 87.2\%]$), 99.6\% fewer copy/paste actions (0.1 vs.\ 29.3; 95\% CI $=[99.1\%, 100.0\%]$), and 80.2\% fewer tab switches (5.8 vs.\ 29.3; 95\% CI $=[75.9\%, 84.1\%]$).
\ai{} required even fewer clicks (10.4 per task) and tab switches (1.4 per task), but shifted substantial manual effort to prompt construction as participants entered an average of 381.6 characters per task with \ai{} (95\% CI $=[307.2, 467.3]$), compared with 10.9 with \sys{} (95\% CI $=[6.1, 16.6]$) and 46.7 with \manual{} (95\% CI $=[26.9, 69.1]$).
% \zh{can potentially talk about the prompting process and thinking about what to prompt led to more incompletions for \ai{}.}

These operational savings translated into shorter task durations, as~\cref{fig:completion-time} shows. Using the 240-s timeout threshold for the sensitivity analysis, the overall participant-level geometric mean duration was 83.0~s with \sys{} (bootstrap 95\% CI $=[67.7, 105.3]$), compared with 171.8~s for \manual{} (95\% CI $=[151.7, 194.4]$) and 177.3~s for \ai{} (95\% CI $=[164.4, 192.8]$). \sys{} therefore required 51.7\% less time than \manual{} (paired-bootstrap 95\% CI $=[41.8\%, 59.3\%]$) and 53.2\% less time than \ai{} (95\% CI $=[40.2\%, 62.0\%]$).
We found a significant overall effect of technique on duration (Friedman $\chi^2(2)=15.17$, $p<0.001$, Kendall's $W=0.63$). Bonferroni-adjusted exact paired Wilcoxon comparisons showed that \sys{} was faster overall than both \manual{} ($p_B=0.001$, $r_{\mathrm{rb}}=1.00$) and \ai{} ($p_B=0.004$, $r_{\mathrm{rb}}=0.95$), while \manual{} and \ai{} did not differ ($p_B=1.000$). At the task level, \sys{} was faster than \manual{} in all three tasks ($p_B=0.015$, $.003$, and $.006$) and faster than \ai{} in Tasks~2 and~3 (both $p_B=0.003$), but not significantly so in Task~1 ($p_B=0.157$). Only 2 of 36 \sys{} trials timed out, compared with 7 of 36 \manual{} trials and 11 of 36 \ai{} trials.

\begin{figure*}[t]
    \centering
    \includegraphics[width=0.72\linewidth]{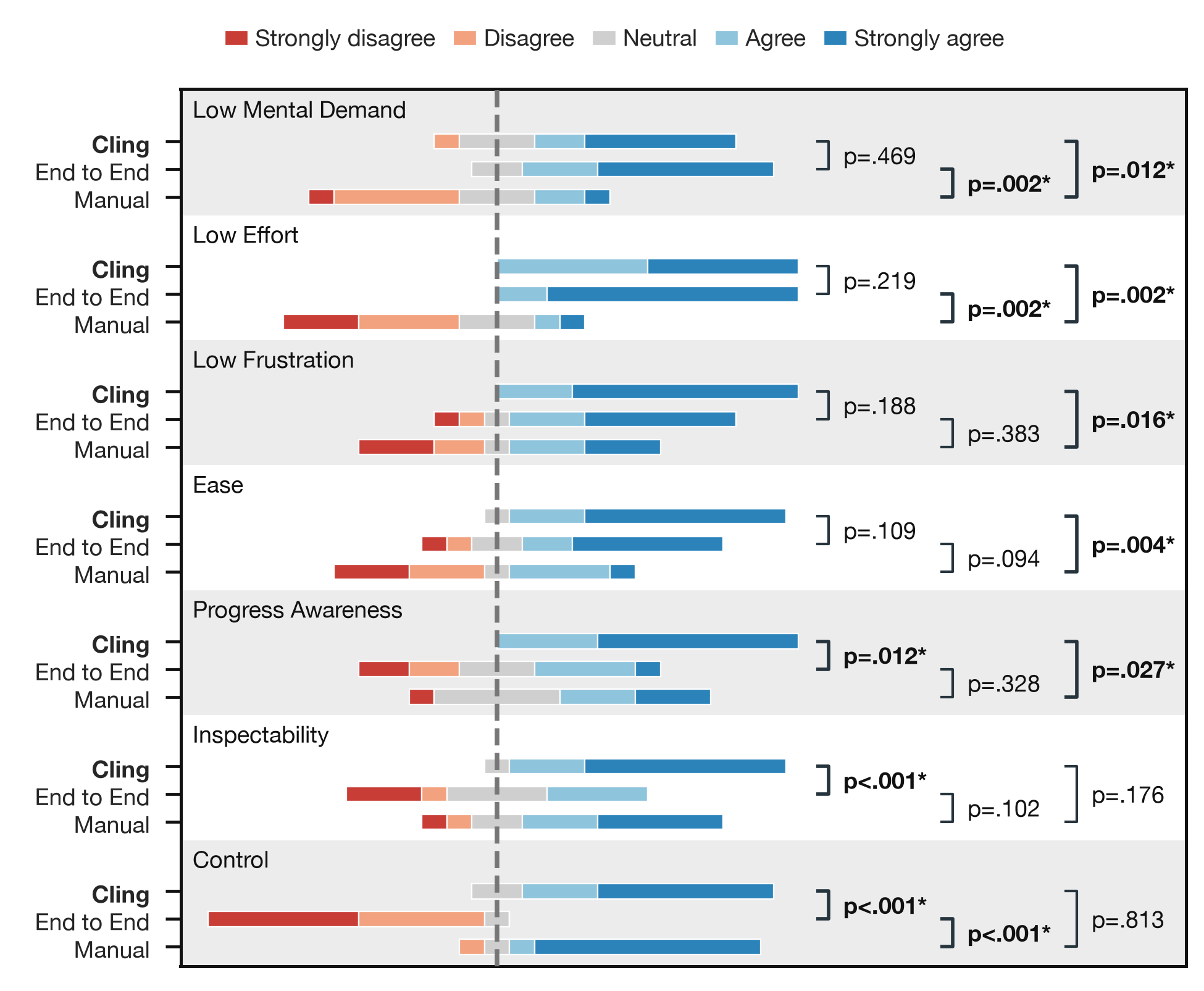}
    \caption{\textbf{Part~1 Questionnaire Ratings.} Both \sys{} and \ai{} received more favorable ratings than \manual{} for mental demand and effort, while \sys{} offered a greater sense of control than \ai{}. Bars show response distributions; brackets report two-sided Wilcoxon signed-rank test $p$-values, with asterisks indicating $p<0.05$.}
    \label{fig:part1_likert}
    \Description[Diverging stacked bar charts]{Diverging stacked bar charts compare Cling, End-to-End, and Manual on seven dimensions: low mental demand, low effort, low frustration, ease, progress awareness, inspectability, and control. Responses range from strongly disagree in red to strongly agree in blue, with neutral responses in gray. Higher agreement indicates more favorable ratings. Cling and End-to-End receive significantly more favorable ratings than Manual for mental demand and effort. Cling also receives significantly higher ratings than Manual for low frustration, ease, and progress awareness, and than End-to-End for progress awareness, inspectability, and control. Manual receives significantly higher control ratings than End-to-End. Other comparisons are not statistically significant.}
\end{figure*}

\heading{\sys{} Enables Low-Effort Yet Controllable Automation}
We compared questionnaire ratings between conditions using two-sided Wilcoxon signed-rank tests, as shown in \cref{fig:part1_likert}.
There is a stark tradeoff of the cost and benefit between \ai{} and \manual{}.
By offloading laborious work to \ai{}, participants perceived lower mental demand
(\ai{} rated more favorably over \manual{} by 1.67 points (95\% CI [1.00, 2.25], $p=0.002$)
and low effort
(\ai{} rated more favorably over \manual{} by 2.42 points, 95\% CI [1.58, 3.17], $p=0.002$),
at the cost of a reduced sense of control
(\ai{} rated lower than \manual{} by 2.92 points, 95\% CI [2.25, 3.50], $p<0.001$).

By allowing participants to automate with small steps and incrementally as needs arose,\sys{} was rated 1.33 points higher than \manual{} for low mental demand (95\% CI [0.58, 2.08], $p=0.012$) and 2.08 points higher for low effort (95\% CI [1.33, 2.83], $p=0.002$).
At the same time, \sys{} was rated 2.83 points higher than \ai{} for control (95\% CI [2.33, 3.33], $p<0.001$).
At the same time, \sys{} was rated 2.83 points higher than \ai{} for control (95\% CI [2.33, 3.33], $p<0.001$).
P9 described \sys{} as a tool they could control, whereas \ai{} felt like ``[it's] working on [its] own.'' They compared \sys{} to a self-retracting tape measure---the button automates retraction, but the user still does the measuring and interpretation.

% Participants rated Cling higher than \ai{} on five of the seven dimensions on average.
% Ranked from largest to smallest differences, Cling increased Control by 2.83 points (95\% CI [2.33, 3.33], $p=0.010$), Inspectability by 1.83 (95\% CI [1.17, 2.58], $p=0.019$), Progress Awareness by 1.67 (95\% CI [0.75, 2.58], $p=0.469$), Low Frustration by 0.75 (95\% CI [0.00, 1.58], $p=1.000$), and Ease by 0.67 (95\% CI [0.00, 1.42], $p=1.000$).
% Cling was rated 0.33 points lower for both Low Mental Demand (95\% CI [$-0.92$, 0.33], $p=1.000$) and Low Effort (95\% CI [$-0.67$, 0.00], $p=1.000$).

% Compared with Manual, Cling was rated higher on six dimensions: Low Effort by 2.08 points (95\% CI [1.33, 2.83], $p=0.035$), Ease by 1.83 (95\% CI [1.00, 2.67], $p=0.059$), Low Frustration by 1.67 (95\% CI [0.75, 2.58], $p=0.219$), Low Mental Demand by 1.33 (95\% CI [0.58, 2.08], $p=0.279$), Progress Awareness by 1.08 (95\% CI [0.25, 1.92], $p=1.000$), and Inspectability by 0.75 (95\% CI [$-0.08$, 1.58], $p=1.000$).
% Control was similar between Cling and Manual (mean difference $=-0.08$, 95\% CI [$-0.75$, 0.67], $p=1.000$).

% All reported $p$-values are Holm-adjusted exact paired sign-test $p$-values across the 21 Part~1 comparisons.
% The 95\% CIs are unadjusted, so CI exclusion of zero does not necessarily correspond to an adjusted $p$-value below $.05$.

\begin{figure*}[t]
    \centering
    \includegraphics[width=0.76\linewidth]{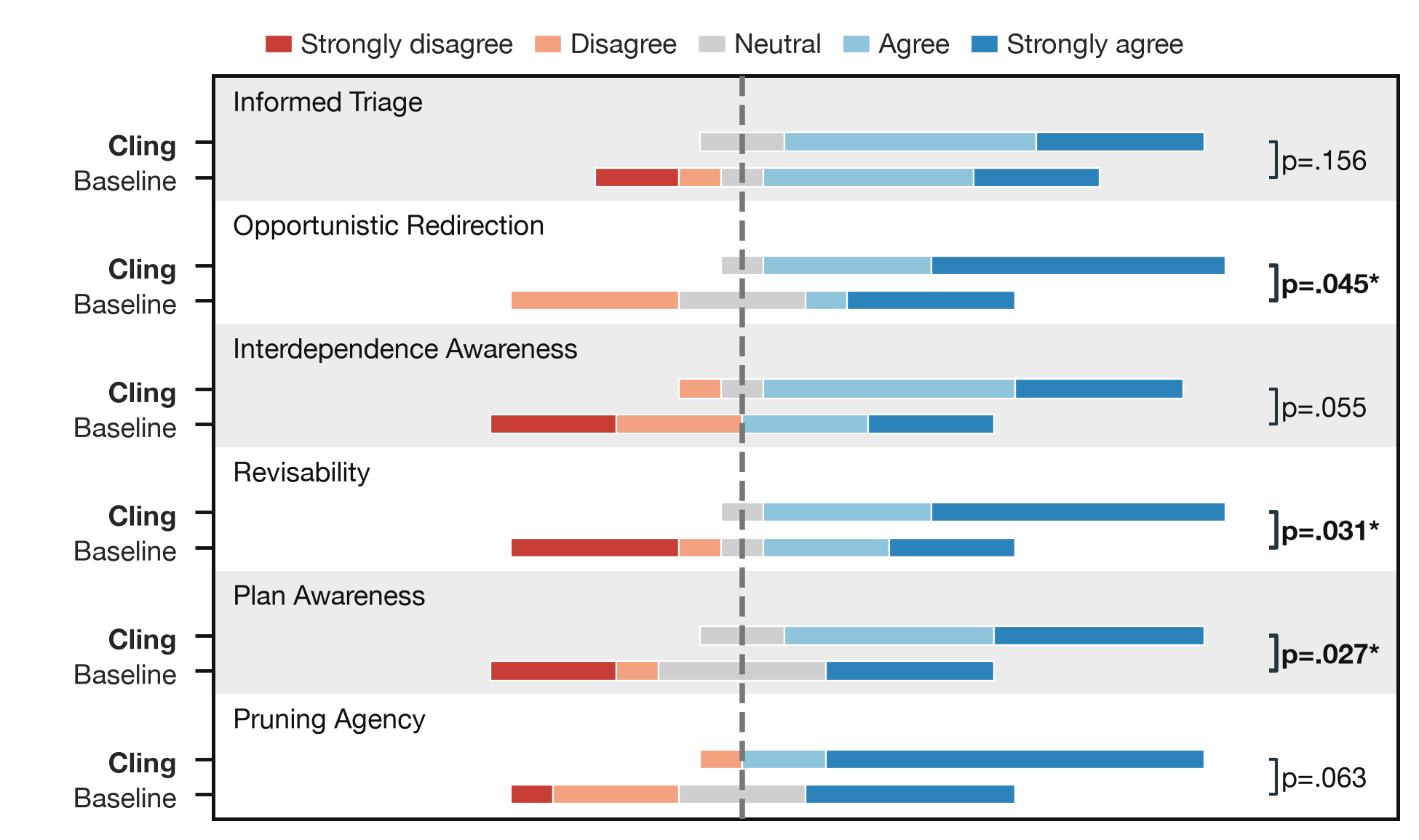}
    \caption{\textbf{Part~2 Questionnaire Ratings.} \sys{} received more favorable ratings than the baseline for opportunistic redirection, revisability, and plan awareness. Bars show response distributions; brackets report two-sided Wilcoxon signed-rank test $p$-values, with asterisks indicating $p<0.05$.}
    \label{fig:part2_likert}
    \Description[Diverging stacked bar charts]{Diverging stacked bar charts compare Cling and the baseline on six dimensions: informed triage, opportunistic redirection, interdependence awareness, revisability, plan awareness, and pruning agency. Responses range from strongly disagree in red to strongly agree in blue, with neutral responses in gray. Cling's responses lean more toward agreement across all six dimensions. Differences favoring Cling are statistically significant for opportunistic redirection, p = .045; revisability, p = .031; and plan awareness, p = .027. Differences are not statistically significant for informed triage, p = .156; interdependence awareness, p = .055; or pruning agency, p = .063.}
\end{figure*}

\heading{In-situ, Small-Step Automation Supports Inspection As Automation Unfolds}
Among completed attempts, precision and recall were high across all three conditions. \manual{} yielded 259 TP, 599 TN, 13 FP, and 3 FN (participant-level mean precision: 94.6\%; recall: 98.9\%). \ai{} yielded 167 TP, 394 TN, 4 FP, and 5 FN (98.3\%; 96.9\%). \sys{} yielded 260 TP, 576 TN, 4 FP, and no FN (97.9\%; 100\%). Friedman tests found no significant effect of condition on precision, $\chi^2(2)=2.70$, $p=0.260$, $W=0.112$, or recall, $\chi^2(2)=2.00$, $p=0.368$, $W=0.083$. We therefore did not conduct the planned pairwise comparisons.

To capture errors that final-artifact analyses would miss, we also examined errors and corrections during task execution.
These execution errors included typos, insertion of entries that did not satisfy the inclusion criteria, and entry of incorrect information.
In \manual{}, 6 of 12 participants made at least one execution error, with 12 errors recorded in total.
Four participants corrected 9 of these errors (75\%).
No such errors were observed in \ai{}.
In \sys{}, 2 of 12 participants made one execution error each, and both errors were corrected.

Participants attributes their ability to catch and correct mistakes to being able to apply small step automation using \sys{} in-situ, similar to manual experience.
Because of \sys{}'s tightly-coupled intent specification and automation, P5 felt they were handling ``input and output at the same time'' and therefore ``[built] trust along the way.'' P8 valued seeing ``where the things are coming from,'' while P2 emphasized that \sys{} navigated directly to the relevant content on the page and made inspection easy.

Inspection of \ai{}, by contrast, involves assessing whether the agent's procedure matched participants' expectations.
P10 described ``using the thinking trace to compare how I would be doing it'' and wanting to inspect further when the procedure diverged from their mental model.
Such mismatches could raise concerns without necessarily indicating an error.
In the paper-transfer task, for example, P3 worried that an eligible paper had been missed when the agent did not immediately enter it into the spreadsheet, although the agent was retaining results for a later batch update.
Verifying \ai{}'s outputs could also require independently repeating parts of the task.
% P11 used \texttt{Command+F} to check one match and one nonmatch against the source page, limiting repeated work at the cost of verification coverage.
P8 questioned the benefit of automation when comprehensive verification would require another manual pass: ``if I am doing that already, then why to use AI?''
% With \sys{}, participants could instead make an additional pass through coordinated items using the stepper, checking results against their sources without repeating the searches manually.

Moreover, the different inspectability influenced participants' comfort level in sharing the resulting artifacts with collaborators as is.
Most participants (P1, P2, P3, P5, P7, P9, P10, P11) placed \sys{} as a close second to \manual{}.
P1 and P2 placed \sys{} above \manual{}, citing concerns about the reliability of manual \texttt{Command+F} searches (P1) and easier inspection through automatic navigation to source content (P2).

\heading{\sys{} Accommodates Intent As It Was Discovered and Evolved Through On-Page Information}
Throughout the Part 2 tasks, people used information on the webpages as evidence for evaluating options and as a resource through which participants formed, refined, and redirected their intent, even when the information was peripheral to the initial task.
P3, for example, described ``discovering [their] taste'' while browsing furniture, including preferences for colors and materials they had not known they had. P7 developed criteria from what was visible on the page: browsing bed-frame images prompted them to consider adding a headboard, while a lamp page's built-in filters reminded them that they did not want a portable lamp because it would require another desk to place it on. Page content also shifted existing priorities: after inspecting roughly ten options, P7 reconsidered an initial emphasis on price and decided that ``money is less of a priority'' than they had expected.
P12 mentioned that seeing different couches could ``jog my memory'' of comfortable furniture at friends' homes, providing visual cues that helped them clarify what they wanted.

In-situ information could also prompt participants to reconsider constraints or redirect their plans.
P12 valued seeing options beyond their current criteria because an appealing \$150 item encountered during a \$100 search, for example, might justify revising the budget rather than simply being filtered away.
P11's plan changed more substantially: they initially intended to find a desk, but encountering a gaming-room category led them to pursue a gaming desk instead.
Reflecting on such changes, P11 described ``being able to improvise'' as important to the exploratory process.

Results from Part 2's questionnaire responses corroborates the theme above (\cref{fig:part2_likert}).
Compared with the baseline, \sys{} was rated 1.17 points higher for plan awareness (95\% CI [0.33, 2.00], $p=0.027$), 1.50 points higher for revisability (95\% CI [0.50, 2.50], $p=0.031$), and 1.08 points higher for opportunistic redirection (95\% CI [0.33, 1.83], $p=0.045$).

To complement the findings above, we analyzed how participants updated their criteria during the open-ended furnishing tasks. We counted criteria articulated during think-aloud or externalized in the artifact, including filtering constraints reused in \sys{} and from prompts sent to the AI assistant in the baseline.
The combined count of criterion additions and updates was higher with \sys{} than with the baseline ($M=6.58$ vs.\ $3.83$; Holm-adjusted $p=0.41$). 
Criteria updates specifically were also more frequent with \sys{} ($M=1.50$ vs.\ $0.08$; two-sided exact paired sign-flip test, Holm-adjusted $p=0.008$).
Although updates were modest in absolute number, eleven of twelve participants updated at least one criterion with \sys{}, compared with only one participant in the baseline.

% \heading{\sys{} Supports Flexible Re-anchoring}

\section{Discussion}
\label{sec:discussion}

\subsection{Situated Automation vs. End-to-End Delegation} 
\sys{} makes automation available as coordination needs arise during ongoing exploration. This is particularly beneficial when tasks involve tacit, dynamic, or compounded constraints that emerge through exploration. For example, when deciding which furniture to buy, P7 reported that only after extended comparison did a small set of options begin to “make sense.” What makes an option desirable may depend on many interdependent considerations and be burdensome to articulate. The situated automation supported by \sys{} does not need to know why an option is desirable; it only needs enough information to carry out the coordination that follows from the user’s choice. This contrasts with task-level delegation in end-to-end automation, which pressures users to translate contextual and still-evolving intent into language. P1 described the “inertia to put things into words,” while P10 preferred directly changing an explicit filter rather than explaining through chat that a previous constraint no longer applied.

Participants also reported that \sys{} allowed them to remain actively involved for enjoyment, learning, or a sense of responsibility, without being burdened by manual work they considered ``noise'' (P5). For example, P7 explained that when exploring an unfamiliar space, they might work more manually or ``offload a lot'' depending on “how much I want to learn.” P2 chose to rely on their own judgment when selecting furniture simply because doing so was ``fun.'' P4 connected the cross-page transfer task to a recurring work activity in which they would shortlist papers from a large pool for a mentee. They felt \sys{} would be a good fit because it could offload the manual foraging of papers while leaving the substantive judgment of suitability to her. 

More broadly, this suggests an automate-as-you-go model in which exploration, automation, and assessment remain
tightly coupled. Rather than requiring users to specify a complete task and then inspect its outcome, automation can be
invoked incrementally as new needs emerge.
Instead of merely remaining present to approve or correct automated actions, the user would exert his or her own judgements and redirect the task as their intent develops.

\subsection{Schema Correspondence as a Basis for Controllable Automation}

In this work, we formalize cross-page coordination as a pervasive interaction pattern in everyday information work and introduce a schema-correspondence perspective that provides a concrete vocabulary for a previously fragmented set of low-level tasks, including copying and pasting, information retrieval, content reformatting, and transfer. Rather than treating these as separate interaction techniques, schema correspondence characterizes them in terms of how entities and attributes on one page relate to those on another.

This abstraction also provides a basis for more controllable automation. LLMs can recover schemas and semantic correspondences from heterogeneous webpages, while explicit rules deterministically specify how information should be selected, aligned, transformed, and transferred. This separation confines probabilistic reasoning to interpreting heterogeneous interfaces while keeping execution predictable and inspectable. Because the resulting rules operate over semantic structures rather than page-specific DOM elements, they can also be inspected, modified, reused, and composed as coordination needs evolve.

\subsection{Toward a Semantically Structured Web}

In the current Web ecosystem and the design of \sys{}, the schema of each page must often be reconstructed on the client side at use time, causing different users to repeatedly infer similar semantic structures from the same webpages. A more efficient architecture would allow webpages to expose their entities, attributes, and relationships directly as machine-readable schemas, bringing the Web closer to the Semantic Web vision~\cite{bernerslee2001semantic}.

The historical difficulty of this vision was partly practical: authoring and maintaining semantic representations imposed additional work on webpage developers and required agreement on representation conventions. LLMs substantially lower this barrier by making it possible to generate and maintain such representations from existing content and interfaces during authoring time. Importantly, this need not require every webpage to share a universal schema. Individual pages could expose their own local semantic structures, while systems such as \sys{} dynamically infer correspondences between them as coordination needs arise. This suggests a path toward semantic interoperability that combines authoring-time semantic structure with use-time semantic alignment.

\section{Conclusion}
\label{sec:conclusion}
To summarize, we conceptualize cross-page coordination as a distinct class of interaction and characterize it through five recurring coordination patterns. We then introduce a technical pipeline that reconstructs reusable webpage schemas and identifies coordination opportunities through schema correspondence, operationalized in our prototype, \sys{}. Our technical evaluations show that this approach can surface accurate and useful coordination opportunities, while findings from the user study show that in-situ, small-step automation enables users to perform coordination tasks efficiently while remaining in control. More broadly, this work points toward an information environment in which independently developed webpages become interoperable at the point of use, allowing users and AI to collaboratively compose information and functionality across page boundaries to accomplish complex information tasks more effectively.

% We conclude by discussing how this approach points toward a future in which information across application and webpage silos becomes more interoperable for end users.

% \input{sections/2-relatedwork}

\bibliographystyle{ACM-Reference-Format}
\bibliography{main, haijun}
% \appendix
% \input{sections/99-appendix}

\end{document}